\documentclass[useAMS,usenatbib]{mn2e}
\usepackage[pdftex]{graphicx}

\usepackage{astro_journal_names}

\defcitealias{2002ApJ...568..475B}{BSI}
\defcitealias{1980MNRAS.191..483L}{LBE}
\defcitealias{2005MNRAS.363.1092A}{A\&S}

\title[Gravothermal collapse of isolated SIDM haloes]{Gravothermal
  collapse of isolated self-interacting dark matter haloes: $N$-body simulation
  versus the fluid model} \author[Jun Koda \& P. R. Shapiro] {Jun
  Koda$^1$\thanks{E-mail: junkoda@physics.utexas.edu.}
  and Paul R. Shapiro$^1$\thanks{E-mail: shapiro@astro.as.utexas.edu} \\ $^1$
  Department of Astronomy, Texas Cosmology Center, the University of Texas
  at Austin, TX 78712, USA}

\pagerange{\pageref{firstpage}--\pageref{lastpage}} \pubyear{2011}

\begin{document}

\label{firstpage}

\maketitle

\begin{abstract}
  Self-Interacting Dark Matter (SIDM) is a collisional form of cold
  dark matter (CDM), originally proposed to solve problems that arose
  when the collisionless CDM theory of structure formation was
  compared with observations of galaxies on small scales. The
  quantitative impact of the proposed elastic collisions on structure
  formation has been estimated previously by Monte Carlo $N$-body
  simulations and by a conducting fluid model, with apparently
  diverging results. To improve this situation, we make direct
  comparisons between new Monte Carlo $N$-body simulations and
  solutions of the conducting fluid model, for isolated
  SIDM haloes of fixed mass. This allows us to separate cleanly the
  effects of gravothermal relaxation from those of continuous mass
  accretion in an expanding background universe. When these two
  methods were previously applied to halo formation with cosmological
  boundary conditions, they disagreed by an order of magnitude about
  the size of the scattering cross section required to solve the
  so-called `cusp-core problem.' We show here, however, that the
  methods agree with each other within $20$ per cent for
  \textit{isolated} haloes. This suggests that the two methods are
  consistent, and that their disagreement for cosmological haloes is
  not caused by a breakdown of their validity.

  The isolated haloes studied here undergo gravothermal collapse. We
  compare the solutions calculated by these two methods for
  gravothermal collapse starting from several initial conditions,
  including the self-similar solution by
  \citet*[][BSI]{2002ApJ...568..475B}, and the Plummer,
  Navarro-Frenk-White and Hernquist profiles. We compare for the case
  in which the collisional mean free path is comparable to, or greater
  than, the size of the halo core.  This allows us to calibrate the
  heat conduction which accounts for the effect of elastic hard-sphere
  scattering in the fluid model. The amount of tuning of the thermal
  conductivity parameters required to bring the two methods into such
  close agreement for isolated haloes, however, is too small to
  explain the discrepancy found previously in the cosmological
  context. We will discuss the origin of that discrepancy in a
  separate paper.
\end{abstract}

\begin{keywords}
  cosmology: theory -- dark matter -- galaxies: kinematics and
  dynamics -- galaxies: haloes -- methods: numerical -- methods:
  $N$-body simulation.
\end{keywords}

\section{Introduction}

In the currently standard cosmological model ($\Lambda$CDM), a flat
universe with a cosmological constant contains collisionless cold dark
matter as its dominant matter component, perturbed by primordial
Gaussian-random-noise density fluctuations. This model has been highly
successful at explaining observations of the background universe and
large-scale structure.  On small scales, however, the distribution of
dark matter in coordinate and velocity space is not fully understood.
$N$-body simulations show that the density profiles of the virialized
regions (`haloes') that form in this collisionless dark matter are
cuspy, such as the \citet*[][NFW]{1997ApJ...490..493N} profile, in
which $\rho \rightarrow r^{-1}$ toward the centre.  Recent
high-resolution simulations show that the inner profile is not exactly
a power law \citep{2004MNRAS.355..794H, 2005MNRAS.364..665D,
  2006AJ....132.2685M, 2008MNRAS.387..536G, 2009MNRAS.398L..21S,
  2010MNRAS.402...21N}, but it diverges nevertheless. On the other
hand, a cored profile (a profile which flattens in the centre) such as
a pseudo-isothermal profile, is favored by observations of dwarf and
low surface brightness (LSB) galaxies \citep{2003MNRAS.340..657D,
  2004MNRAS.351..903G, 2006ApJS..165..461K,2006A&A...452..857Z,
  2007A&A...467..925G, 2008ApJ...676..920K, 2008AJ....136.2648D,
  2010arXiv1011.0899O, 2010AdAst2010E...5D}. Dwarf spiral and LSB
galaxies are dark-matter dominated. As such, it was originally
thought, their mass distribution should reflect the dark matter
dynamics alone, and be relatively less affected by the complexity of
the dissipative baryonic component. This, it was thought, makes these
systems ideal for studying the undisturbed, intrinsic, dark matter
distribution on small scales.

This apparent cusp-core conflict was one of the small-scale structure
problems of the CDM model which prompted suggestions that the dark
matter might be something else, with microscopic properties that would
alter the structure on small scales without spoiling the success of
CDM on large-scales.  \cite{2000PhRvL..84.3760S} proposed
self-interacting dark matter (SIDM) as a possible solution to this
cusp-core problem, adding hypothetical elastic-scattering collisions
to the otherwise collisionless particles of the standard CDM
cosmology. Heat transfer within the virialized haloes, due to these
non-gravitational collisions, was then suggested to make the halo
cores expand. This latter idea was confirmed by several numerical and
analytical studies.  \citet{2000ApJ...534L.143B} introduced a Monte
Carlo scattering algorithm between dark matter particles to take the
self interaction into account in a numerical $N$-body simulation, and
this method was refined by \citet{2000ApJ...543..514K}. These $N$-body
results suggested, however, that SIDM haloes would undergo gravothermal
collapse, making them unsuitable to explain the observed haloes.

\citet*[hereafter BSI]{2002ApJ...568..475B} applied a conducting fluid model,
originally invented to describe gravitational scattering in star
clusters, to isolated\footnote{The term `isolated halo' shall refer
  here to an object of fixed mass with vacuum boundary conditions;
  i.e., not subject to cosmological boundary conditions involving mass
  infall or evolution by perturbation growth in a
  Friedmann-Robertson-Walker universe.} SIDM haloes.
\citetalias{2002ApJ...568..475B} derived a self-similar gravothermal
collapse solution at large Knudsen number ($\mathit{Kn} \gg 1$, where
$\mathit{Kn}$ is the ratio of SIDM scattering mean free path to the
system size). The isolated halo collapsed within a finite time. They
also showed that the collapse is delayed compared to their
self-similar solution when the Knudsen number is comparable to or
smaller than one, because the length scale of energy exchange is
restricted by the mean free path. Their conclusion that the collapse
time would naturally exceed a Hubble time was more optimistic about
the SIDM hypothesis.

A realistic halo is not isolated, however, since it forms
in a cosmologically expanding background universe, with infall and a
finite pressure at the virial radius \citep{1999MNRAS.307..203S}.
Cosmological simulations showed that a cross section per unit mass
$\sigma = 0.5-5 \textrm{ cm}^2 \textrm{ g}^{-1}$ makes the profile
cored, and the cored profile is stable \citep{2000ApJ...544L..87Y,
  2001ApJ...547..574D}. \citet{2002ApJ...581..777C} emphasized that
the profile depends on the accretion history, especially when the last
major merger occurred.

With the importance of cosmological infall in mind,
\citet{2005MNRAS.363.1092A} derived a cosmological similarity solution
for this problem, with proper account taken of such cosmological
boundary conditions. This solution shows that the gravothermal
collapse, which occurs for the isolated halo, is prevented by infall
in the cosmological case, and the core has a constant size in units of
the virial radius, for a given SIDM cross section. When there is no
self-interaction, this fluid approximation gives a density profile
similar to the cuspy profile found in $N$-body simulations. This shows
that the fluid approximation also describes the virialization of
collisionless dark matter appropriately, providing in effect an
analytical derivation of the NFW profile in the collisionless
limit. This is because the collisionless Boltzmann equation reduces to
fluid conservation equations for an ideal gas with ratio of specific
heats equal to $5/3$ if the velocity distribution of the particles in
the frame of bulk motion is isotropic and skewless (see Section
\ref{sec:basic-equations}).  In the presence of SIDM scattering
collisions, too, those analytical solutions are in qualitative
agreement with the corresponding Monte Carlo $N$-body simulations;
SIDM haloes have cores and the cores collapse within a finite time
when they are isolated, but they do not collapse within a Hubble time
in a cosmological environment. However, the values of the cross
section necessary to explain the observed dark matter density profiles
are not in agreement.  \citet{2005MNRAS.363.1092A} find that $\sigma
\sim 200 \,\textrm{ cm}^2 \textrm{ g}^{-1}$ fits the dwarf and LSB
galaxy rotation curves best, while $N$-body simulations suggest the
range of $\sigma = 0.5-5 \,\textrm{ cm}^2 \mathrm{ g}^{-1}$ gives the
observed central density.

In order to test the SIDM hypothesis by comparison with astronomical
observations, it is necessary to improve our understanding of these
theoretical predictions and reconcile them. That will be the focus of
this paper, as described below. In the meantime, related progress
continues to be made on other fronts, in comparing both SIDM and
collisionless CDM with observation.  For example, non-circular motions
may affect the density profile estimate \citep{2006MNRAS.373.1117H},
but they are usually not strong enough to make observations consistent
with the theoretically-predicted cuspy profile
\citep{2008AJ....136.2761O, 2008AJ....136.2720T, 2009ApJ...692.1321K}.
The cusp-core conflict may be mitigated by baryonic processes of gas
outflow, induced by supernovae feedback \citep{2010Natur.463..203G,
  2010arXiv1011.2777O}. Central dark-matter densities could also be
reduced, in addition, by bars \citep{2002ApJ...580..627W}, or by gas
clumps sinking via dynamical friction \citep{2001ApJ...560..636E},
while the system is baryon rich, but these two scenarios may not be
strong enough to make large enough cores in realistic situations
\citep{2008ApJ...679..379S, 2009ApJ...691.1300J}.  While observations
of dwarf and LSB galaxies generally prefer cored density profiles,
some, however, are also consistent with an NFW profile
\citep{2004MNRAS.355..794H, 2005ApJ...621..757S, 2006MNRAS.373.1117H,
  2007ApJ...657..773V}.  The wide diversity of dwarf/LSB cores has
also led to a suggestion that SIDM alone cannot be the full solution
to the cusp-core problem \citep{2005ApJ...631..244S,
  2010ApJ...710L.161K}.

There are also some constraints on the value of the SIDM cross section
from comparison of theoretical predictions with galaxy clusters, for
which dark matter velocity is much higher than for galaxies.  $N$-body
results find that the core size of relaxed SIDM clusters of galaxies
becomes too large for cross section $\sigma \ga 1 \textrm{ cm}^2
\mathrm{ g}^{-1}$ \citep{2000ApJ...544L..87Y, 2002ApJ...572...66A,
  2003ApJ...586..135L}. The analytical cosmological self-similar
solutions of \citet{2005MNRAS.363.1092A}, however, point out that the
core sizes of clusters are also small enough, not only for
\textit{small} cross section in the long-mean-free-path regime, but
also for \textit{large} cross section in the short-mean-free-math
regime, i.e., $\sigma \ga 200 \textrm{ cm}^2 \mathrm{ g}^{-1}$, because
the short mean free path then limits the amount of heat
conduction. However, such a large cross section would also enhance the
fluid-like behaviour of SIDM, which may then conflict with
observations of merging clusters.  For example, observations of
cluster 1E 0657-56, known as the `bullet cluster,' from which total
matter density has been mapped by weak and strong gravitational
lensing measurements while the density of the intergalactic
baryon-electron fluid was mapped by X-ray measurements, show that the
dark matter spatially segregated from the baryon-electron plasma, as
it would be if the dark matter is not highly collisional
\citep{2006ApJ...648L.109C}. The dark matter and galaxies of the
subcluster have passed through the main cluster without distortion
while the baryon gas shows a bow shock due to its collisional
nature. This observation excludes the possibility that SIDM is too
highly collisional, or fluid-like. Analytical estimates and Monte
Carlo $N$-body simulations of the bullet cluster by
\citet{2004ApJ...606..819M} and \citet{2008ApJ...679.1173R} constrain
the velocity independent cross section to be $\sigma < 0.7 \,\textrm{
  cm}^2 \textrm{ g}^{-1}$, using the fact that the mass-to-light ratio
of the subcluster is normal.  Larger cross section would make the
mass-to-light ratio smaller because the SIDM collisions scatter the
dark matter out of the subhalo.  [On the other hand, see
\citet{2007ApJ...668..806M} for a possibility that SIDM with cross
section $\sim 4 \,\textrm{ cm}^2 \mathrm{ g}^{-1}$ explains a cluster
Abell 520 which has substructures with anomalous mass-to-light
ratio]. If the cross section is velocity \textit{dependent}, it 
puts a constraint only at very large relative velocity. The relative
velocity of the merging haloes is estimated to be between $2,500 $ and
$4,000 \textrm{km s}^{-1}$ at their observed separation, and even
higher at centre passage \citep{2007ApJ...661L.131M,
  2007MNRAS.380..911S, 2008MNRAS.389..967M}.  If the cross section is
too large, the SIDM halo could also be too spherical compared to the
observed elliptical matter distribution of galaxy clusters
\citep{2002ApJ...564...60M}. In addition, large cross sections result
in too much heating and evaporation of substructure haloes in clusters
\citep{2001ApJ...561...61G}.  These constraints almost exclude the
possibility that a velocity-\textit{independent} cross section solves
the cusp-core problem for dwarf/LSB galaxies, but,
velocity-\textit{dependent} cross section can still be effective on
dwarf scales, and simultaneously negligible on galaxy cluster scales:
e.g., inversely proportional to relative speed
\citep{2000astro.ph.10497F, 2003MNRAS.338..156D}, or short range
Yukawa interaction, whose scattering cross section has $v^{-4}$
velocity dependence for large relative velocities
\citep{junkoda.phdthesis, 2010arXiv1011.6374L}.

There was an additional motivation for the SIDM hypothesis when it was
first put forth, involving the overabundance of subhaloes in CDM
$N$-body simulation compared to observations of the Local Group. The
number of substructures in the collisionless CDM model has previously
been thought to be about an order of magnitude larger than the
observed number of dwarf galaxies in the Local Group. Although the
number of substructures is reduced by SIDM stripping,
\citet{2003ApJ...586...12D} claim that SIDM does not solve this
problem; the cross-section that makes the profile cored
($0.6\,\mathrm{cm}^2\mathrm{g}^{-1}$) is apparently not efficient
enough to reduce the number of satellites down to the observed level.
Recent findings of nearby faint galaxies improve the agreement between
observations and collisionless CDM simulations, but still require some
feedback or reionization effect that suppresses star formation in
low-mass haloes \citep[e.g.][and references
therein]{2009ApJ...696.2179K, 2010MNRAS.tmp.1538W}.

Apart from the original motivation for proposing SIDM as a solution of
small-scale structure problem of the collisionless CDM model, interest
in the SIDM hypothesis remains strong for other reasons, as well. The
fundamental nature of dark matter is still unknown.  Anomalous cosmic
ray detections, reported by PAMELA, FERMI and ATIC, have recently
stimulated new particle physics models for dark matter that result in
the scattering interaction of SIDM as a secondary effect
\citep{2009PhRvD..79a5014A, 2010PhRvD..81h3522B, 2010PhRvL.104o1301F}.
Astronomical constraints on the SIDM cross section and its velocity
dependence continue to be of importance, therefore, in order to
constrain such models.

Further progress along these lines requires that we reduce the
uncertainties in the theoretical productions for SIDM. As described
above, the quantitative estimates by Monte Carlo $N$-body simulation
and the conducting fluid model, of cross section values that are
consistent with observations of dwarf galaxy rotation curves, are in
strong disagreement with each other.  As long as this disagreement is
unresolved, many of the \textit{additional} constraints on the SIDM
hypothesis described above, which also depend upon the validity of the
$N$-body results or related analysis, will remain suspect.  It is
possible that either the $N$-body or the fluid model does not describe
the system correctly. To remove such possibility, we directly compare
the two methods in the simplest case, namely, the isolated
spherically symmetric halo.

We will summarize the basis for the conducting fluid model in Section
\ref{section-fluid}. In Section \ref{section-method}, we will describe
our Monte Carlo numerical algorithm for the SIDM elastic scattering
interaction in $N$-body simulations. We test those two methods against
each other for isolated haloes in Section \ref{section-results}. The
impact of our comparison results on the cosmological similarity
solution by \citet{2005MNRAS.363.1092A} is discussed in Section
\ref{section-cosmological-solution}. Finally, our results are
summarized in Section \ref{section-conclusion}.

\section{The Conducting Fluid Model}
\label{section-fluid}

\subsection{Background: Gravothermal Collapse in Star Clusters}
The conducting fluid model (also known as the gaseous model) was first
developed to study gravothermal collapse in globular star cluster
systems, and has been shown to be successful in those
systems. \citet[hereafter LBE]{1980MNRAS.191..483L} proposed a thermal
conduction formula for collisional, gravitationally-bound systems
based on dimensional analysis, and derived an analytical solution that
describes the self-similar collapse of a star cluster.  That
self-similar solution appears in the late stage of collapse in a
Fokker-Planck calculation \citep{1980ApJ...242..765C} and in $N$-body
simulations \citep{1996ApJ...471..796M, 1996MNRAS.282...19S,
  2003MNRAS.341..247B, 2005PhRvL..95h1102S}.  When the time evolution
of a Plummer sphere is solved numerically by integrating the partial
differential equations of the fluid model, the resulting collapse time
agrees with that from other methods \citep{1987ApJ...313..576G,
  1989MNRAS.237..757H}. SIDM haloes and globular clusters are both
`collisional,' self-gravitating systems, but the angular distribution
and velocity dependence of the collisions are different in the two
cases. Stars obey Rutherford scattering, which is dominated by
small-angle scattering and small velocity encounters, $\sigma \propto
v^{-4}$, while the SIDM cross section we explore in this paper is
isotropic and velocity independent. It is possible, however, that the
SIDM interaction also obeys Rutherford scattering via a `dark-photon'
interaction \citep{2009PhRvD..79b3519A,
  2009JCAP...07..004F}. Gravitational Rutherford scattering between
dark matter particles are negligible unless they
are all $10^5 - 10^6 M_{\sun}$ mass black holes
\citep{2005MNRAS.359..104J}.

\subsection{The Basic Equations}
\label{sec:basic-equations}
In this section, we review the conducting fluid model developed by
\citetalias{1980MNRAS.191..483L} and \citetalias{2002ApJ...568..475B}
before we compare its results with our simulations. As in these
papers, we shall here restrict the conducting model to the case of
particle distributions that are spherically symmetric, isotropic in
velocity dispersion, and quasi-static. The quasi-static approximation
means that, while the fluid evolves thermally, it always satisfies
hydrostatic equilibrium at each moment. For the problem at hand, it is
a good approximation, because the collapse timescale is always much
longer than the dynamical time.  The conducting fluid model is not, in
general, restricted to quasi-static systems, however
\citep{1983MNRAS.203..811B, 2005MNRAS.363.1092A}.  Deviation from
isotropic velocity dispersion is also possible to consider, but it
plays only a minor role for the problem of interest here.

The Boltzmann equation, which is a partial differential equation in
phase space, can be written as an infinite series of moment equations
in position space by integrating over all velocities. When third
moments are negligible (skewless), the series of moment equations can
be closed by truncating at second order, as described in
\citet{2005MNRAS.363.1092A}. The equations which result in this case
are what we shall here refer to as `the fluid approximation.' The same
equations would result from assuming a Gaussian velocity distribution
(not necessarily isotropic or Maxwellian), an approximation called
`Gaussian closure' \citep[][and references therein]
{1996JSP....83.1021L}. If velocity isotropy is imposed in addition,
the fluid approximation gives the familiar Euler equation for an ideal
gas with the ratio of specific heats, $\gamma = 5/3$.  This fluid
approximation can describe the structure formation of collisionless
CDM, and accurately reproduce the CDM halo properties found in 3D
$N$-body simulations when applied to the problem of spherical
cosmological infall \citep{2005MNRAS.363.1092A}.

When collisions are important (e.g. either gravitational scattering between
stars or DM self-interaction), thermal conduction, which is a third
order moment, should not be neglected. However, accurate evaluation of
third moments is only successful in the small Knudsen number regime,
$\mathit{Kn} \ll 1$. In this regime, the Navier-Stakes equation can be
derived from the Boltzmann equation with the Fourier law of heat flux,
\begin{equation}
\label{eq-fourier-law}
  \frac{L_\mathrm{smfp}}{4\pi r^2} = -\frac{3}{2} a^{-1} b \rho
  \frac{\lambda^2}{t_r} \frac{\partial v^2}{\partial r},
\end{equation}
where $\rho(r,t)$ is the mass density, $v(r,t)$ is the one-dimensional
velocity dispersion, $\sigma$ is the scattering cross section
\textit{per unit mass}, $\lambda \equiv 1/\rho \sigma $ is the mean
free path, and $ b \equiv 25 \sqrt{\pi}/32 \approx 1.38$ is a
constant\footnote{BSI used a value $ b = 25 \sqrt{\pi} / (32 \sqrt{6})
  \approx 1.002 $, but it should be $ 25 \pi /32$.}
\citep[Chapman-Enskog theory, e.g.,][]{ 1970mtnu.book.....C,
  Lifshitz}.  The local relaxation time is defined as $t_r(r,t) \equiv
1/(a\rho\sigma v)$, where the constant $a \equiv \sqrt{16/\pi} \approx
2.26 $ describes the collision rate of particles that follow a
Maxwellian distribution, defined in \citetalias{2002ApJ...568..475B}.

In the other limit, $\mathit{Kn} \gg 1$, Lynden-Bell \& Eggleton
\citepalias{1980MNRAS.191..483L} found that an empirical thermal conduction
formula with $\lambda$ replaced by the gravitational scale height (or
Jeans length), $ H \equiv \sqrt{v^2/ 4 \pi G \rho} $, explains the
gravothermal catastrophe of star clusters very well, i.e., 
\begin{equation}
\label{eq-lbe}
  \frac{L_\mathrm{lmfp}}{4\pi r^2} = -\frac{3}{2} C \rho
  \frac{H^2}{t_r} \frac{\partial v^2}{\partial r},
\end{equation}
where $C$ is an unknown constant of order unity. The scale height $H$
characterizes the length scale that particles (or stars) orbit under
the action of the gravitational force.

BSI combined the two limiting forms of this thermal conduction into one as follows:
\begin{equation}
  \label{eq-bsi_heat}
  \frac{L}{4 \pi r^2} = -\frac{3}{2} \rho
    \left[
      \left( C \frac{H^2}{t_r} \right)^{-1} +
      \left( a^{-1} b \frac{\lambda^2}{t_r} \right)^{-1}
    \right]^{-1}
    \frac{\partial v^2}{\partial r}.
\end{equation}
The first term inside the brackets is the LBE formula, equation (\ref{eq-lbe}),
which dominates in the large Knudsen number limit, $\lambda \gg H$.
In the other limit ($\lambda \ll H$), the second term inside the brackets
dominates, and the conduction converges to
equation~(\ref{eq-fourier-law}), instead.  BSI's formula is an
empirical interpolation between those two heat conductivity limits.  BSI
assumed $ C=b $, but the exact value cannot be determined
analytically. We determine the value of $C$ by fitting the Monte Carlo
$N$-body data in Section \ref{section-bsi}.

The conducting fluid model is a set of moment equations closed empirically
by this conductive heat flux $L/4\pi r^2$, as follows:
\begin{eqnarray}
  \label{eq-fluid}
  \frac{\partial}{\partial r} (\rho v^2) &=& -\rho \frac{GM}{r^2},
  \\ \label{eq-fluid2} -\frac{1}{4\pi r^2} \frac{\partial L}{\partial r} &=& \rho v^2
  \frac{D}{Dt} \ln \left( \frac{v^3}{\rho} \right),
\end{eqnarray}
where $M$ is the mass enclosed by radius $r$, and $D/Dt$ is the
Lagrangian derivative with respect to time. The first equation describes
hydrostatic equilibrium. The second equation is the first law of
thermodynamics, in the form relating energy transfer and entropy,
`$dQ=TdS=dE+pdV$.'

When $\lambda \gg H$, the fluid equations (\ref{eq-fluid},
\ref{eq-fluid2}) with the heat conduction equation (\ref{eq-lbe}) have
a self-similar solution.  In general, if a self-similar solution
exists, if density is static as $r \rightarrow \infty$ and if the
evolution timescale $ \rho_c/ \dot{\rho}_c $ is proportional to the
relaxation time at the centre, $ t_{r,c}(t) \equiv t_r(r=0, t) $, then
the central quantities evolve as,
\begin{equation}
  \label{eq-self_similar_1}
  \rho_c(t) / \rho_c(0) =
    \left( 1 - t/t_\mathit{coll} \right)^{-2 \alpha/( 3 \alpha -2) },
\end{equation}
\begin{equation}
  \label{eq-self_similar_2}
  v_c^2(t) / v_c^2(0) = 
    \left( 1 - t/t_\mathit{coll} \right)^{ -(2 \alpha - 2)/(3 \alpha - 2)},
\end{equation}
\begin{equation}
  \label{eq-self_similar_3}
  t_{r,c}(t)/t_{r,c}(0) = 1 - t/t_\mathit{coll},
\end{equation}
for some constants $ \alpha $ and $ t_\mathit{coll} $, by dimensional
analysis \citepalias{1980MNRAS.191..483L}. The exponents of
$1-t/t_\mathit{coll}$ depend on the form of the relaxation time,
therefore, different from those for star clusters.  
BSI obtained,
\begin{equation}
  \label{eq-self_similar_alpha}
  \alpha= 2.190,
\end{equation}
\begin{equation}
  \label{eq-self_similar_tcoll}
  t_\mathit{coll} = 290 \,C^{-1} t_{r,c}(0),
\end{equation}
by solving an eigenvalue problem of a system of ordinary differential
equations.
 
For the non-self-similar time evolution considered in Sections
\ref{section-plummer}, \ref{section-singular}, and
\ref{section-transitional}, we must integrate the time evolution of
fluid variables $\rho$ and $v^2$ numerically by alternative steps of
the heat conduction and adiabatic relaxation to hydrostatic
equilibrium, as described in \citetalias{2002ApJ...568..475B}.

\section{$N$-body Method with Monte Carlo Scattering}
\label{section-method}

\subsection{Scattering Algorithm}
In this section, we will describe the Monte Carlo algorithm we
implemented to model non-gravitational scattering of dark matter
particles by other dark matter particles, within a pre-existing
gravitational $N$-body method. Our scattering algorithm is similar to
\citet{2000ApJ...543..514K}. Each particle can collide with one of its
$k$ nearest neighbors with a probability consistent with a given
scattering cross section.  For simplicity, we assume collisions are
elastic, velocity independent, and isotropic in the centre of mass
frame, but the Monte Carlo method can handle any differential cross
section. We first outline the Monte Carlo $N$-body method and then
explain, in detail, the algorithm that we have implemented in the
parallel $N$-body code \textsc{GADGET 1.1}
\citep*{2001NewA....6...79S}, which uses the tree algorithm to
calculate gravitational forces.

Monte Carlo algorithms for particle-particle scattering (known as
direct simulation Monte Carlo) have been used for more than thirty
years to solve physics and engineering problems of collisional
molecules, giving reasonable results \citep{bird}. For example, the
results agree with an exact solution of the spatially homogeneous
Boltzmann equation that describes the relaxation toward a Maxwellian
distribution; they also agree with the Navier-Stokes equation solutions
and experiments, including the thermal conductivity, in the small
$\mathit{Kn}$ regime \citep[e.g.,][]{1984PhFl...27.2632N, bird,
  PhysRevE.69.042201}.

Consider $N$-body particles at positions $\textbf{x}_j$ and velocities
$\textbf{v}_j$ with equal mass $m$. We discretize the distribution
function $f$ with,
\begin{equation}
  f(\textbf{x}, \textbf{v}) = \sum_j m W(\textbf{x}-\textbf{x}_j;
  r_j^{k\mathrm{th}}) \delta^3(\textbf{v}-\textbf{v}_j),
\end{equation}
where $W(\textbf{x}; r_k)$ is a spline kernel function
of size $r_k$, $r_j^{k\mathrm{th}}$ is the distance
from particle $j$ to its $k\approx 32$nd nearest neighbor, and
$\delta$ is the Dirac delta function. Our choice of kernel is often
used in smoothed particle hydrodynamics, including \textsc{GADGET}.
Our algorithm is identical to that of Kochanek \& White if a top-hat
kernel is used for $W$ instead of a spline. The result does not depend
on the details of the kernel, however. We tested with $k=128$ but did
not see any difference.

The collision rate $\Gamma$ for a particle at position $\textbf{x}$ with
velocity $\textbf{v}$ to collide with this distribution $f$ is,
\begin{equation}
  \Gamma = \sum_j m W(\textbf{x}-\textbf{x}_j; r_j^{k\mathrm{th}}) \sigma
  |\textbf{v}-\textbf{v}_j|,
\end{equation}
where $\sigma$ is the scattering cross section per unit mass.
Therefore the probability that an $N$-body particle $0$ collides with
particle $j$ during a small timestep $\Delta t$ is,
\begin{equation}
  P_{0j} = m W(\textbf{x}_0-\textbf{x}_j; r_j^{k\mathrm{th}}) \sigma
 |\textbf{v}_0-\textbf{v}_j| \Delta t
  \label{eq-pairwise-probability}
\end{equation}

One can generate a random number and decide whether this collision
happens and reorient velocities when they collide. This method is
similar to a variant of direct simulation Monte Carlo called Nanbu's
method \citep{1980JPSJ...49.2042N}. His Monte Carlo algorithm, with
the pairwise collision probability,
equation~(\ref{eq-pairwise-probability}), can be derived from the Boltzmann
equation as described in his paper. Conversely, results of Nanbu's
numerical method converge, mathematically, to the solution of the
Boltzmann equation as the number of particles goes to infinity
(\citealt{babovsky}).  In Nanbu's method, only one particle is
scattered per collision (only particle $0$ but not $j$). The
philosophy is that the $N$-body particles are samples chosen from real
sets of microscopic particles, and those samples should collide with
a smooth underlying distribution function, not necessarily with
another sampled $N$-body particle.  However, then the energy and
momentum are not conserved per collision. Moreover, the expectation
value of the energy decreases systematically \citep{greengard}. In our
case, the error in the energy rises by $10$ per cent quickly, so we decided to
scatter $N$-body particles in pairs, not using Nanbu's
method. Scattering in pairs is common in direct simulation Monte Carlo
(e.g. Bird's method).

When particles are scattered in pairs, other particles $j$ can scatter
particle $0$ during their timestep as well, but
the scattering probability $P_{0j}$, in equation
(\ref{eq-pairwise-probability}), is similar to, but not exactly equal
to $P_{j0}$, due to the difference in kernel sizes. Therefore, we
symmetrize the scattering probability by taking the average scattering
rate. Note that it is not trivial to generalize the pairwise
scattering algorithm to simulations with unequal $N$-body particle
masses, because $P_{0j}$ and $P_{j0}$ would then differ by a factor of
their mass ratio; there is no reason to symmetrize two intrinsically
different probabilities into single pairwise scattering probability.

In the following, we describe our algorithm in detail. Each particle,
say particle $0$, goes through the following steps, (i) to (iii),
during its timestep $\Delta t_0$. Let particles $1, \dots, k$ be the
$k$ nearest neighbors of particle $0$ ($k=32 \pm 2$). The particle $0$
collides with its neighbors with probabilities $P_{j0}/2$ (equation
\ref{eq-pairwise-probability}) during its timestep. The factor of two
is the symmetrization factor that corrects the double counting of
pairs. A particle $j$ would also scatter particle $0$ during its
timestep, which results in a symmetrized scattering rate. Imagine a
probability space $[0,1]$, with disjoint subsets $I_j \equiv
[\sum_{l=1}^{j-1} P_{l0}/2, \sum_{l=1}^{j} P_{l0}/2)$ that represent
scattering events between particles $0$ and $j$. We neglect the
possibility of multiple scattering in the given timestep. Particle $0$
collides with at most one of its neighbors. We restrict the timestep
so that it is small enough that this approximation is good enough (see
equation \ref{eq-condition_P} below). We generate a uniform random number
$x$ in $[0,1]$, and scatter particles $0$ and $j$ if $x$ falls in a
segment $I_j$, as described below.

\begin{enumerate}
\item In the large $\mathit{Kn}$ regime, most of the particles do not
  collide with another particle in a given timestep. Therefore, we can
  reduce the computation by estimating the rough scattering
  probability first, and compute the accurate probability $P_{0j}$
  only if necessary. First, we calculated an upper bound to the
  scattering probability,
  \begin{equation}
    \label{eq-sup-P}
    \bar{P} = \tilde{\rho} \sigma v_\mathit{max} \Delta t_0,
  \end{equation}
  where $\tilde{\rho}$ is the approximate density calculated from
  $r_0^{k\mathrm{th}}$ via $\tilde{\rho} = km/\frac{4}{3} \pi
  (r_0^{k\mathrm{th}})^3$, and $v_\mathit{max}$ is the maximum speed
  of all the particles. If the generated random number $x$ is larger
  than $\bar{P}$, this means that $x$ is not in any segment $I_j$,
  therefore, the particle $0$ does not collide during this timestep.

\item If the possibility of collision was not rejected in step (i), we
  calculate the pairwise scattering probability $P_{j0}$ and determine
  which neighbor the particle $0$ collides with.  The index $j$ of the
  collision partner is the smallest integer that satisfies $x \le
  \sum_{l=1}^j P_{0j}$, i.e. $x \in I_j$, if such $j$ exists
  (otherwise, the particle does not collide with any neighbors).

\item For particle pairs that collide, we reorient their velocities
  randomly, assuming an elastic scattering which is isotropic in the
  centre of mass frame. Isotropic random directions can be generated
  by one square root operation, without using trigonometric functions
  \citep[e.g.,][]{Vesely}. The velocities are updated in the kick
  phase of the leap-frog time integration in the \textsc{GADGET 1.1}
  gravitational $N$-body method. At that time, we also update the
  centre-of-mass velocities of the nodes around the scattered
  particles in the oct-tree, used for the gravity calculation. This is
  because the centre-of-mass velocities can be changed drastically by
  scattering.
\end{enumerate}

We allowed at most one scattering per timestep per particle.  In
order to suppress the error due to possible multiple scattering, we
restrict the timestep so that,
\begin{equation}
  \label{eq-condition_P}
  \bar{P} < 0.1.
\end{equation}
This restriction makes the Monte Carlo method computationally costly
in the small $\mathit{Kn}$ regime, because then the timestep becomes
much smaller than the dynamical time -- of the order of the timestep
in collisionless $N$-body simulations. This timestep problem may seem
to be avoided by performing multiple scatterings per dynamical
timestep, but there is another limit when the Knudsen number is small.
The distances to $k$th neighbors must be smaller than the mean free
path $\lambda=1/(\rho \sigma)$,
\begin{equation}
\label{eq-condition_r32}
  r_k \la \lambda.
\end{equation}
Otherwise, particles more than a mean free path away would be allowed
to collide and, thereby, make the heat transfer larger than it should
be in the diffusion limit.  If one tries to avoid this by choosing a
kernel size smaller than the mean particle separation length, then the
particles simply freely stream beyond the mean free path, which is
again incorrect. The only way to overcome this problem is to increase
the number of the particles in inverse proportion to the volume within
the mean free path, $N_\mathrm{particles} \propto \lambda^{-3} = (\rho
\sigma)^3$, which increases very rapidly during core collapse as
$\rho$ increases. Conditions (\ref{eq-condition_P}) and
(\ref{eq-condition_r32}) prevent us from running simulations with
small mean free path.

\subsection{Dimensions and Units}
\label{section-definition}
We use the following characteristic scales in the rest of this paper.
For the initial BSI self-similar profile (Sections \ref{section-bsi}, 
\ref{section-transitional}) and the Plummer model (Section
\ref{section-plummer}), we describe densities and velocities in units
of central quantities, $\rho_c(t) \equiv \rho(0,t)$ and $v^2_c(t)
\equiv v^2(0,t)$; $\rho(r,t)$ and $v(r,t)$ denote the density and the
one dimensional velocity dispersion, respectively, in spherical
symmetry. We use core radius $r_c(t) \equiv v_c/\sqrt{4\pi G\rho_c}$
as a standard length scale.

For Hernquist and NFW profiles (Section~\ref{section-singular}), which
have singularities at the centre, we use the scale radius $r_s$ and
density $\rho_0$ that appear in the density profiles as units,
instead. We use $v_0 \equiv r_s \sqrt{4 \pi G \rho_0}$ as the velocity
scale, which is similar to the definition of the core radius above.

The relaxation times at the centre $t_{r,c}(t) \equiv t_r(0,t)$ for
BSI and Plummer model, or $t_{r,0} \equiv 1/(a\rho_0\sigma v_0) $ for
NFW and Hernquist profiles, are used as unit time-scales.

We express cross sections in a dimensionless way as $\hat{\sigma}_c
\equiv \rho_c \sigma r_c$ and $\hat{\sigma}_0 \equiv \rho_0 \sigma
r_s$, where $\sigma$ is the scattering cross section per unit
mass. The inverses are the Knudsen numbers, the mean free path in the
unit of system size ($r_c$ or $r_s$). The evolution of the system is
characterized by the Knudsen number $\mathit{Kn}$ only, not depending
on the overall physical scale.

\subsection{Simulation Setup}
We generate the initial conditions for $N$-body particle positions and
velocities randomly from the distribution functions using the
rejection method \citep{1974A&A....37..183A}.\footnote{See also
  \textit{The Art of Computational Science} vol. 11 by Hut, P. and
  Makino,
  J.;\\ \texttt{http://www.artcompsci.org/kali/vol/plummer/title.html}}
The distribution function of the BSI's self-similar solution and the
NFW profile are calculated numerically by Eddington's formula
\citep{1987gady.book.....B}.  The distribution functions of the
Plummer model and the Hernquist model \citep{1990ApJ...356..359H} have
known analytical forms.  We set the initial centre-of-mass velocity of
the system of $N$-body particles to zero by an overall boost.

\begin{table*}
  \begin{tabular}{llllrrrl}
    \hline
    run name & initial condition & cross section & $ \phantom{ab} N$ &  $N_c^{\max} $ & $r_{f}  \quad$     & $\epsilon\quad$ & $\> C$ \\
    \hline
    BSI & BSI self-similar  & $\hat{\sigma}_c = 0.067$   & $4 \times 64^3$ & $473$       &  $600 \,r_c$   & $0.1 \,r_c$ & 0.75 \\
    P & Plummer           & $\hat{\sigma}_c = 0.067$   & $4 \times 32^3$ & $1455$      &  $58.5 \,r_c$  & $0.1 \,r_c$ & 0.8 \\
    H & Hernquist profile & $\hat{\sigma}_0 = 0.16$  & $2 \times 64^3$ & $1493$      &  $100 \,r_s$   & $0.03 \,r_s$ & 0.9 \\
    NFW & NFW profile       & $\hat{\sigma}_0 = 0.088$ & $2 \times 64^3$ &  $794$      &  $100 \,r_s$   & $0.03 \,r_s$ & 0.75 \\ 
    \hline
  \end{tabular}
  \caption{The long-mean-free-path regime: parameters used for each
    run. $N$ is the number of $N$-body particles, $ N_c^{\max} $ is the
    maximum number of particles inside the core, $r_f$ is the radius of
    the reflecting boundary, $\epsilon$ is the initial Plummer 
    equivalent gravitational softening length, which is reset sometimes 
    overtime
    as central density $\rho_c(t)$ increases; i.e. $\epsilon \propto
    r_c(t) \propto v_c^2/\sqrt{\rho_c}$. The constant $C$ is the LBE prefactor
    used for the conducting fluid model.}
  \label{table-parameters}
\end{table*}

We truncate the initial profile at some radius $r_f$, and put a simple
reflecting boundary, which flips the direction of the radial velocity
if particles are moving outward passed the reflecting boundary. We use
$r_f = 600 \,r_c$ for the BSI self-similar profile, $r_f = 58.5 \,r_c$
for the Plummer model, and $r_f = 100 \,r_s$ for the Hernquist and NFW
profiles.  The density at $r_f$ is smaller than $2\times10^{-7}
\rho_c$, and the heat flux ($L/4\pi r^2$ in Section
\ref{section-fluid}) at $r_f$, calculated from the equation for the
conducting fluid model, is smaller than $0.02$ per cent of its maximum
value.  Nevertheless, the collapse time is sometimes sensitive to the
position of this reflecting boundary. The collapse was slower by a
factor of two when we first used truncation radius $r_f= 58.5 \,r_c$
for the BSI profile, even though this $r_f$ is large enough to satisfy
Antonov's criterion for gravothermal catastrophe
\citep{1997PASJ...49..345E}. We also tested $r_f= 300 \,r_c$, for the
BSI profile, and the collapse time changed by only $3$ per cent
compared to $r_f=600 \,r_c$. Hence, our choice of $r_f = 600 \,r_c$ is
large enough to bring about a converged solution to high accuracy.

We use two timestep criteria in addition to equation
(\ref{eq-condition_P}). The timestep must also satisfy $\Delta t \le
\eta_v v_c(0)/a$, and $\Delta t \le \eta_G / \sqrt{G\tilde{\rho}}$,
where $v_c(0)$ is the initial one-dimensional velocity dispersion, $a$
is the local acceleration, and $\tilde{\rho}$ is the local density
calculated from $k=32$ nearest neighbors (see below equation
\ref{eq-sup-P}). We choose the dimensionless parameters to be $\eta_v
= 0.02$ and $\eta_G = 0.005$.  With this choice, the initial
conditions are static for several dynamical times when scattering is
turned off. Energy conservation is satisfied to better than $1$ per
cent in all runs.

The number of particles, gravitational softening length and other
parameters are summarized in Table \ref{table-parameters}.  For cases
BSI and P, we reset the gravitational softening length $\epsilon$ to
$0.1 \,r_c(t)$ every time the central density increases by a factor of
10, to avoid the numerical effect of softening on the density profile.

We tested that the numerical scattering rate is correct, by counting
the number of scatterings in the simulation of a non-singular
isothermal sphere\footnote{The solution of hydrostatic equilibrium
  (equation \ref{eq-fluid}) with constant $v(r) \equiv v_c$ and finite
  central density.} and comparing it to the analytical rate. The
scattering rates agree within $3$ per cent when $64^3$ particles are
used and the isothermal sphere is truncated at $58.5 \,r_c$. The
difference is due to the fluctuation in the randomly generated initial
condition, not to the Poisson fluctuation in the number of
scatterings.

\subsection{Analysis Methods}
We calculate the central quantities for each snapshot from $N$-body
particles within a sphere of radius $r_c$ around the density-weighted
centre of mass \citep{1985ApJ...298...80C}, assuming an isothermal
sphere inside $r_c$. Namely, we find a consistent $r_c$ iteratively
that satisfies,
\begin{equation}
  r_c = \sqrt{v_c^2/4 \pi G \rho_c},
\end{equation}
where $\rho_c$ is the central density estimated from $M(r_c)$, the
mass inside $r_c$,
\begin{equation}
  \label{eq:rhoc_analysis}
  \rho_c \equiv 1.10 \times M(r_c)/\frac{4}{3} \pi r_c^3,
\end{equation}
and $v_c$ is the velocity dispersion inside $r_c$. The value $1.10$ in
equation~(\ref{eq:rhoc_analysis}) is the ratio of the central density
to the average density inside $r_c$ for the non-singular isothermal
sphere.  In this way, we can use as many particles as possible without
systematic error for the calculation of the central quantities.  The
velocity dispersion inside the core quickly becomes isothermal due to
collisions.

We calculated the smoothed density and velocity dispersion field at
each point in space using an adaptive kernel, similar to that used in the
well-known SPH method, with a Gaussian kernel whose size is $r_{32}$ --
the distance to the 32nd nearest neighbor. These smoothed-particle
density and velocity dispersion fields were then averaged over
spherical surfaces on different radii $r$ to give the radial profiles
of these quantities. In practice, this amounts to summing the
spherically-averaged Gaussian kernels evaluated at each radius $r$
\citep[e.g.,][]{2005MNRAS.357...82R}.

\section{Results}
\label{section-results}
We compare our Monte Carlo $N$-body simulations with the conducting
fluid model, first in the large $\mathit{Kn}$ regime for various
initial conditions, and then in the transitional regime $\mathit{Kn}
\sim 1$.

\subsection{Long Mean-Free-Path Regime}
\subsubsection{Self-Similar Gravothermal Collapse Solution}
\label{section-bsi}
\begin{figure}
\includegraphics[width=84mm]{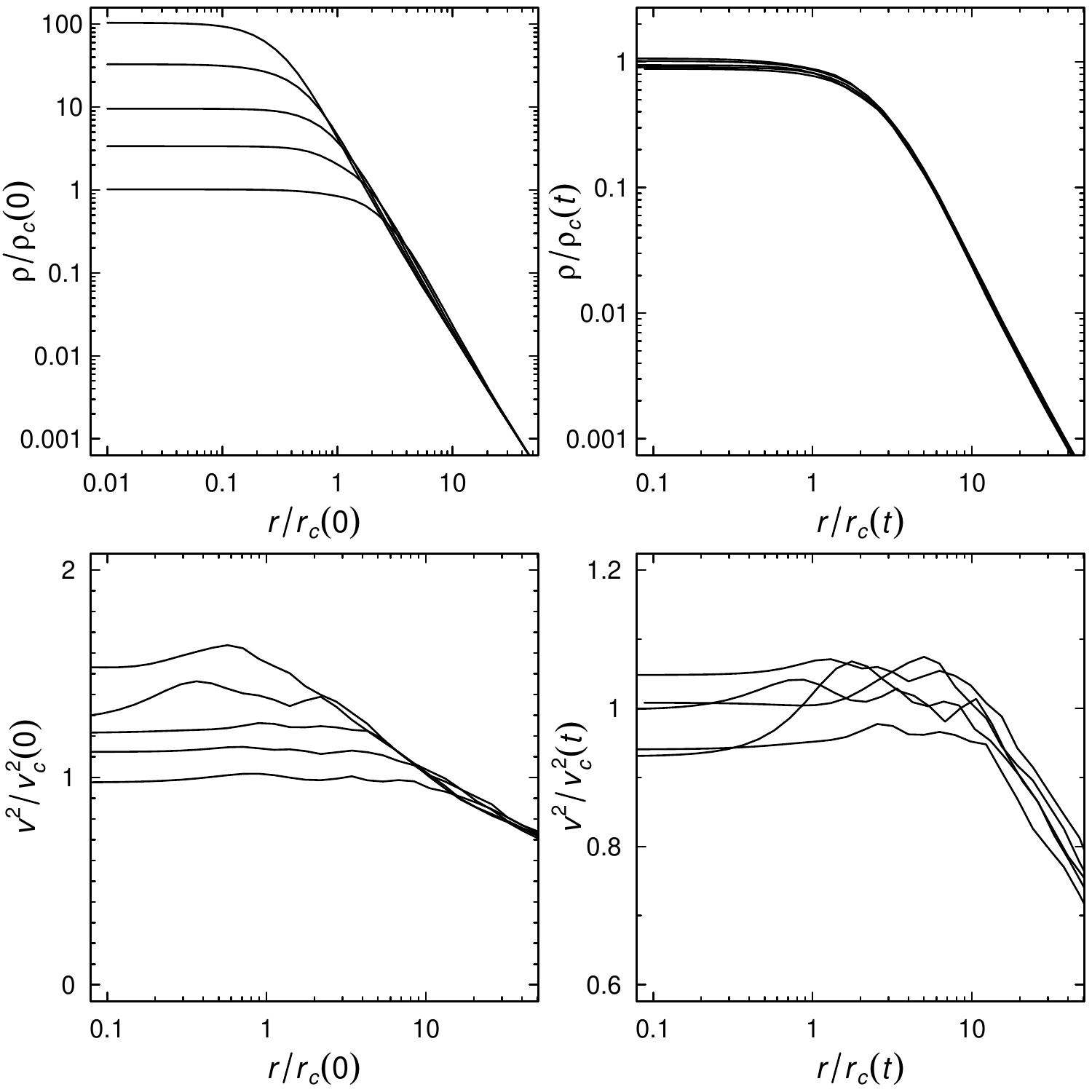}
\caption{Monte Carlo $N$-body results for evolution from BSI
  self-similar initial conditions. Profiles of density (\textit{top})
  and velocity dispersion (\textit{bottom}) for $\hat{\sigma}_c =
  0.067$ plotted in units of fixed, initial central values, as
  labelled (\textit{left}), and in units of the time varying
  self-similar quantities (\textit{right}) at $t/t_{r,c}(0)= 0, 275,
  356, 376$ and $383$. The evolution is indeed self-similar.}
\label{fig-bsi_profile}
\end{figure}

In this section, our Monte Carlo $N$-body simulation for large
$\mathit{Kn}$, $ \hat{\sigma}_c(0) = \mathit{Kn}^{-1}= 0.067 $, is
compared with the BSI self-similar solution.
Fig.~\ref{fig-bsi_profile} shows the evolution of the density and
velocity dispersion profiles. The right panels, plotted in
self-similar variables, show that, when the $N$-body particles are
initialized according to a given time-slice of the self-similar
solution, they indeed evolve self-similarly in the Monte Carlo
$N$-body simulation, thereafter.

\begin{figure}
\includegraphics[width=84mm]{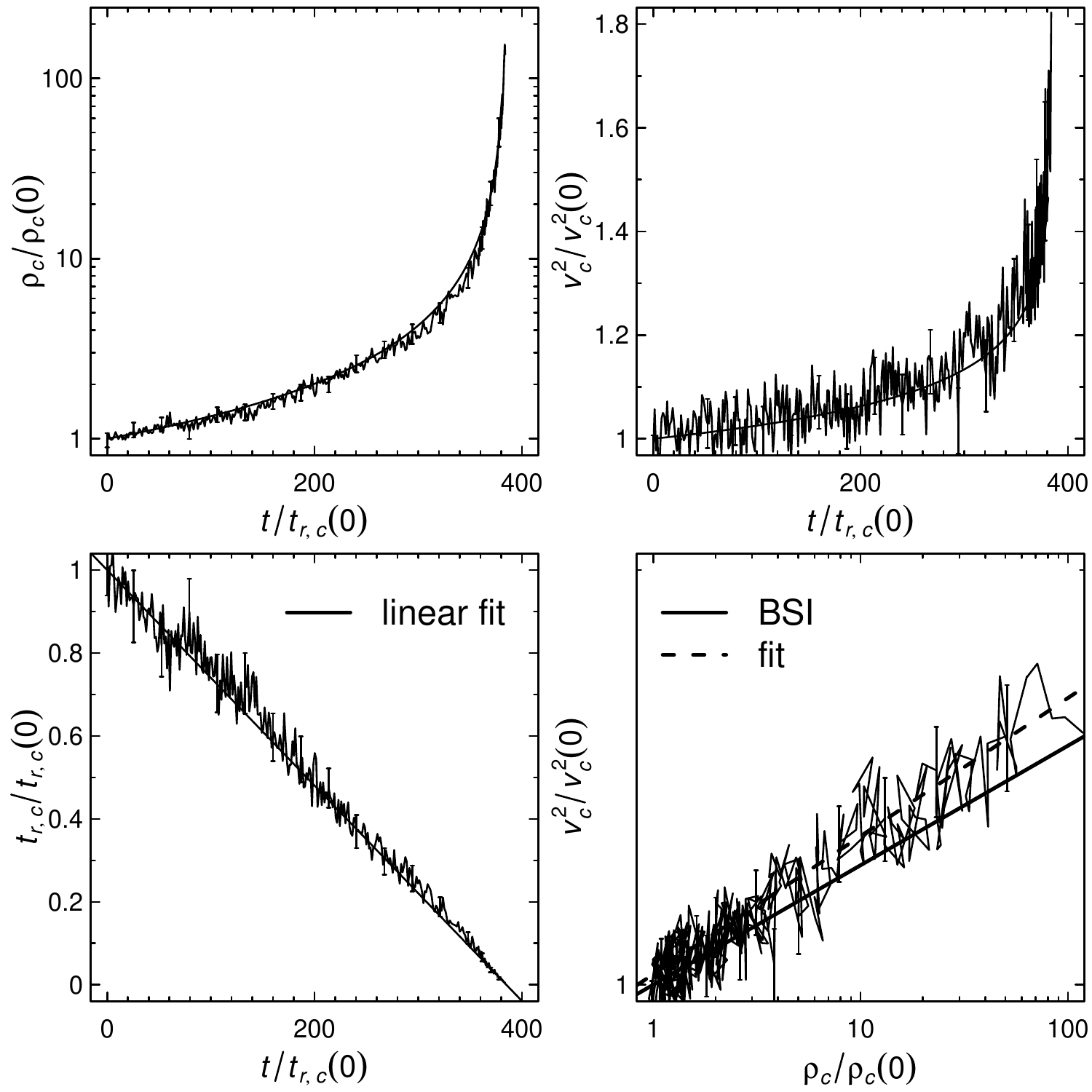}
\caption{Comparison of $N$-body results and BSI similarity solution
  for the case shown in Fig.~\ref{fig-bsi_profile}. Central quantities
  plotted as a function time. $N$-body results are in good agreement
  with the self-similar solution (smooth curves, equations
  \ref{eq-self_similar_1}--\ref{eq-self_similar_alpha}), with
  $t_\mathit{coll}= 385 \,t_\mathit{r,c}(0)$ or $C=0.75$.
  Fluctuations are of order Poisson noise; error bars represent $
  \Delta \rho_c/\rho_c = \Delta t_{r,c}/ t_{r,c} = 2/\sqrt{N_c} $, and
  $\Delta v_c^2/v_c^2 = 1/\sqrt{N_c}$, plotted for every 10 data
  points, where $N_c$ is the number of particles inside $r_c$.  }
\label{fig-bsi_central}
\end{figure}

Fig. \ref{fig-bsi_central} shows that our Monte Carlo $N$-body
simulation is in excellent agreement with the self-similar solution
(equations \ref{eq-self_similar_1}-\ref{eq-self_similar_tcoll}), by
adjusting the value of one parameter, $C$, in the heat conduction
equation (\ref{eq-lbe}). We determine the collapse time
$t_\mathit{coll}= 385\, t_{r,c}(0)$ by fitting the time evolution of
the relaxation time data (Fig. \ref{fig-bsi_central}, left-bottom
panel) by the linear function, equation~(\ref{eq-self_similar_3}). The
prefactor of the thermal flux $C = 0.75 $ follows from
equation~(\ref{eq-self_similar_tcoll}).  The best-fitting power law
index of $ v_c^2 \propto \rho_c^{(\alpha-2)/\alpha} $ (equations
\ref{eq-self_similar_1} and \ref{eq-self_similar_2}) gives a value
$\alpha=2.22$ (Fig.~\ref{fig-bsi_central}, right-bottom), which agrees
reasonably well with the value of the self-similar solution, $2.19$
(equation \ref{eq-self_similar_alpha}).

\begin{figure}
\includegraphics[width=84mm]{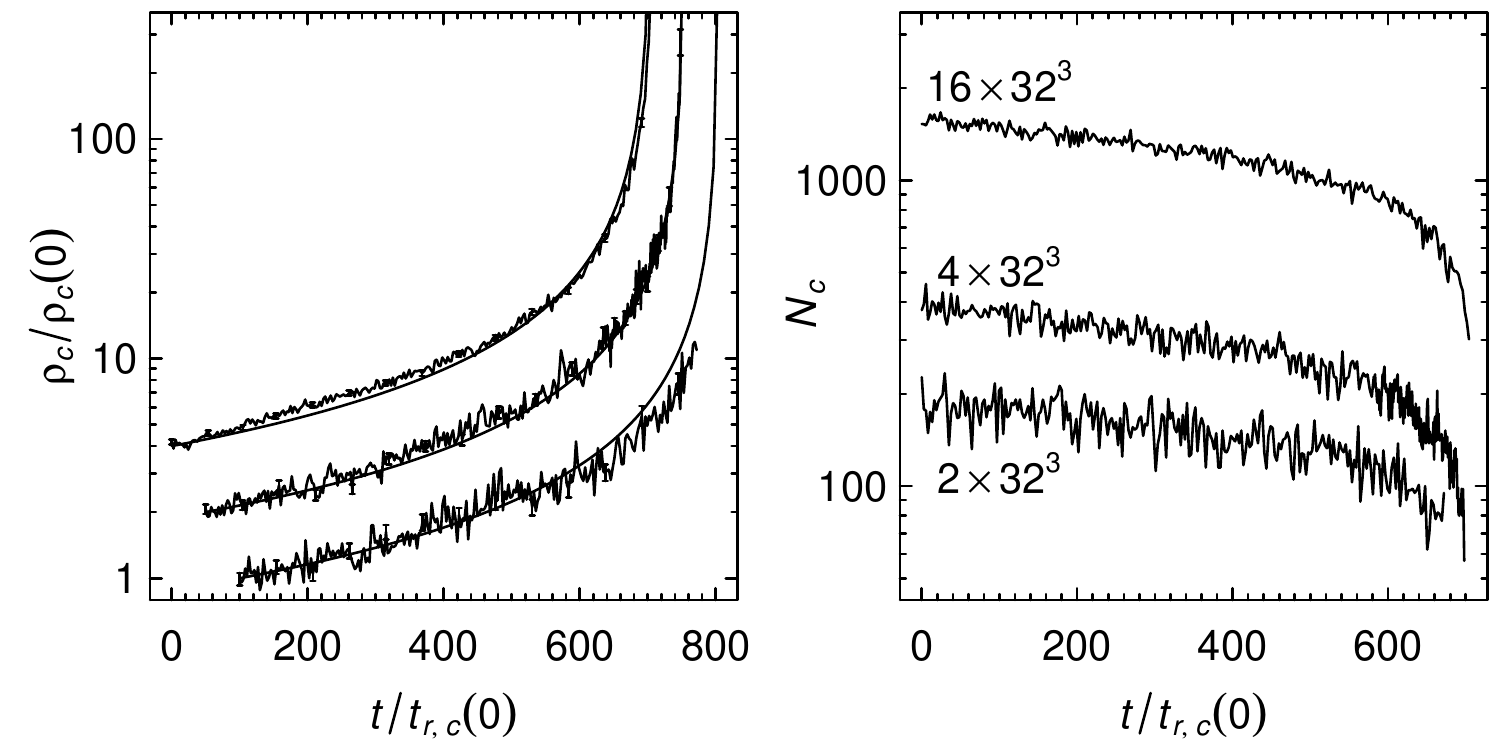}
\caption{Same as Fig.~\ref{fig-bsi_central}, except for simulations
  with different total numbers of $N$-body particles. Density deviates
  from BSI solution when $N_c \la 100$.  Collapse times are different
  from those in run BSI because the reflecting boundary is set at
  $58.5 \,r_c$. Density curves (left panel) are shifted left by 50,
  and up by factor of 2 for readability; three smooth curves are all
  the same BSI self-similar solution. }
\label{fig-n-test}
\end{figure}

To test convergence, we simulated the self-similar solution with three
different numbers of particles, $2\times 32^3$, $4\times 32^3$, and
$16\times 32^3$, which give the initial number of particles inside the
core $N_c(0)= 227, 376$ and $1527$, respectively. At the time of our
resolution test, we used reflecting radius $r_f = 58.5 \,r_c$.
Numerical errors due to finite number of particles should only depend
on the $N$-body particle density, independent of the choice of
boundary condition. As a result, it was not necessary to run
additional convergence test for the case with $r_f = 600 \,r_c$, used in
our final runs. Other parameters are the same as run BSI. In
Fig. \ref{fig-n-test}, we plot the evolution of density and number of
particles inside the core as a function of time.  The three smooth
curves in the left panel are the same analytic self-similar solution
(equation~\ref{eq-self_similar_1}) with $t_\mathit{coll}= 705
\,t_{r,c}(0)$.  The results for the runs with $N= 4 \times 32^3$ and
$N= 16\times 32^3$ agree very well, while those for $N= 2 \times 32^3$
deviate from the other two runs systematically when $N_c \la 100$. Our
run BSI contains 470 particles inside the core at $t=0$
(Table~\ref{table-parameters}), which is better than the converged
$N=4 \times 32^3$ run here. Other runs in the following sections have
better resolution inside the core, because of the smaller fraction of
particles for $r > r_c$, resulting from the steeper decline in their
density profiles.

\subsubsection{Plummer Model}
\label{section-plummer}
We compare the Monte Carlo $N$-body simulation with the conducting
fluid model when the initial condition is the Plummer model in the
large $\mathit{Kn}$ regime. The Plummer model, a standard initial
condition used to study gravothermal instabilities, has a spherical
mass distribution given by,
\begin{equation}
  \rho(r) = \frac{M_T}{4\pi a_{pl}^3/3} \frac{1}{(1+r/a_{pl})^{5/2}}.
\end{equation}
In this case, the characteristic scales defined in
Section~\ref{section-definition} can be shown to be, $v_c =
(2GM_T/a_{pl})^{1/2}/12$ and $ r_c= a_{pl}/(3 \sqrt{2})$. We evolve
this system according to the conducting fluid model in the large
$\mathit{Kn}$ limit with the quasi-static approximation described in
Section \ref{section-fluid}, equations (\ref{eq-lbe}),
(\ref{eq-fluid}), and (\ref{eq-fluid2}).  The $N$-body simulation is
also performed in the large $\mathit{Kn}$ regime, $\hat{\sigma}_c(0)=
0.013$. The time evolution for this quasi-static system is independent
of the actual value of $\sigma$ for the large $\mathit{Kn}$ regime, if
the time for solutions with different $\sigma$ is expressed in units
of the relaxation time.

\begin{figure}
\includegraphics[width=84mm]{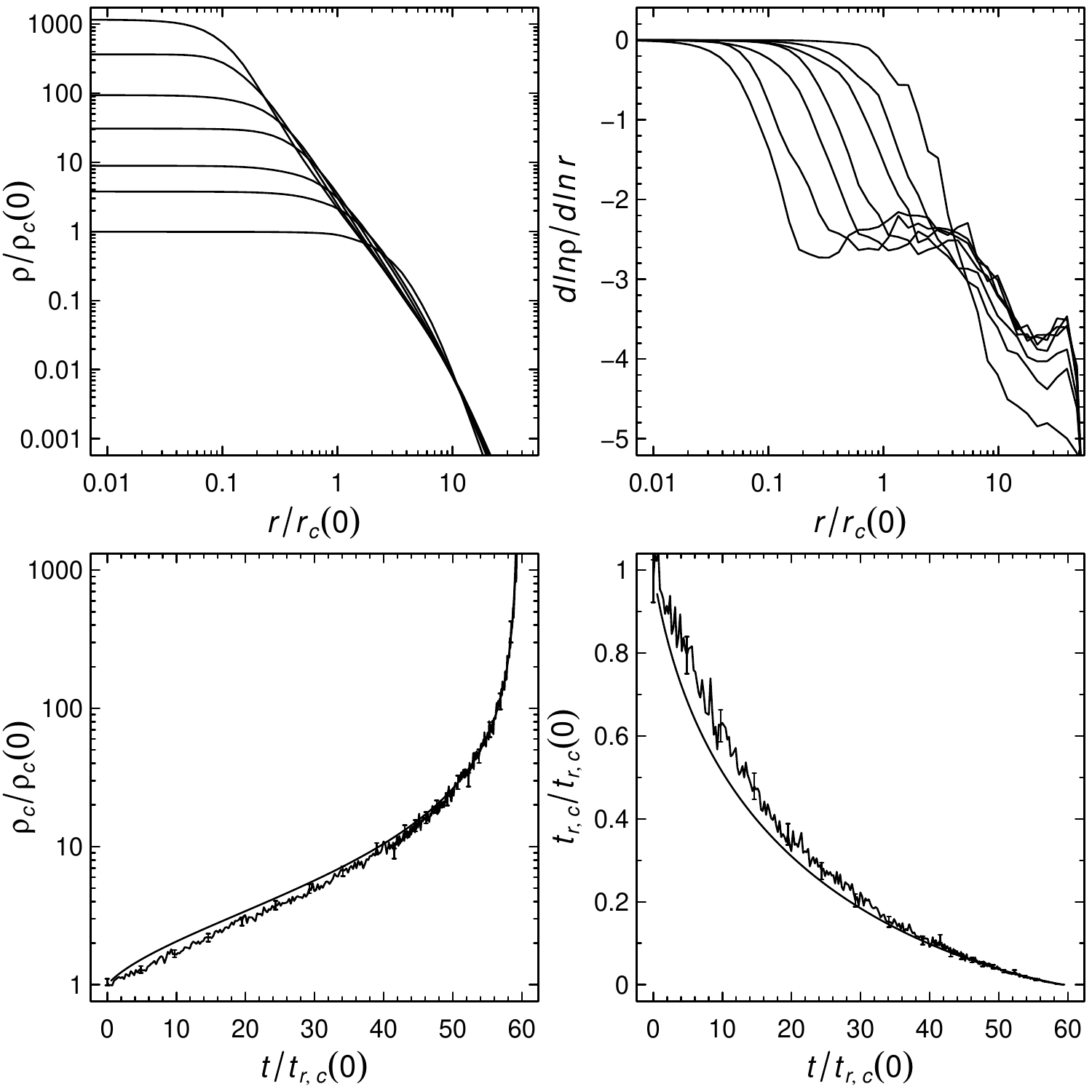}
\caption{Collapse of the Plummer model for $\hat{\sigma}_c(0)=
  0.013$. (\textit{Top:}) Snapshots of the Monte Carlo $N$-body
  simulation, taken at $t/t_{r,c}(0)= 0.0, 24.3 38.5 52.0, 56.7,
  58.6,$ and $59.3$.  (\textit{Bottom:}) Central density and
  relaxation time evolution as functions of time for both the
  $N$-body results and the 1D calculation of the fluid model with
  $C=0.8$ (smooth curves). }
\label{fig-plummer}
\end{figure}

We plot the time evolution in Fig.~\ref{fig-plummer}. The fluid model
agrees with the $N$-body results reasonably well if the coefficient
$C$ is assigned a value of $0.80$.  This is in reasonably good
agreement with the value of $C=0.75$, found for the self-similar
solution. The logarithmic slope (right-top panel) has a plateau at
about $-\alpha$, which is the asymptotic slope of the self-similar
profile. This implies that the inner part is converging to the
self-similar solution, with an asymptotic logarithmic slope $-\alpha$,
which was well known in the gravothermal collapse of star clusters.

\subsubsection{NFW \& Hernquist Profiles}
\label{section-singular}
We also considered the case of an initial condition with spherical
mass distribution given by the NFW profile \citep{1997ApJ...490..493N}
to test the consistency of the Monte Carlo $N$-body simulation and the
conducting fluid model with each other for the typical halo profile
seen in cosmological $N$-body simulations.  The gravothermal collapse
timescale is interesting as a way to put constraints on the SIDM cross
section, because the result of gravothermal collapse is a divergent
profile with $\rho \propto r^{-2.2}$, which is not seen in
observations. We note that gravothermal collapse is prevented by
cosmological infall or major mergers, but when a halo decouples from
cosmological growth, the halo could experience gravothermal collapse.
The NFW profile is given by,
\begin{equation}
  \rho(r) = \rho_0 \frac{1}{r/r_s (1+r/r_s)^2}.
\end{equation}

In addition, we perform a test with initial conditions based upon a
 Hernquist profile \citep{1990ApJ...356..359H},
\begin{equation}
\rho(r) = \rho_0 \frac{1}{r/r_s (1+r/r_s)^3},
\end{equation}
which has the same inner profile, $\rho \propto r^{-1}$, and is
sometimes used to approximate the NFW profile. As we will demonstrate
below, the gravothermal collapses for the NFW and Hernquist profiles
are significantly different, and it is therefore, dangerous to use the
results for the Hernquist profile \citep{2000ApJ...543..514K} to
describe realistic haloes.  As for the Plummer model, our conducting
fluid model calculations adopt the large $\mathit{Kn}$ limit, and the
$N$-body simulations use a small, but nonzero, $\sigma$
(cf. Table~\ref{table-parameters}).

\begin{figure}
\includegraphics[width=84mm]{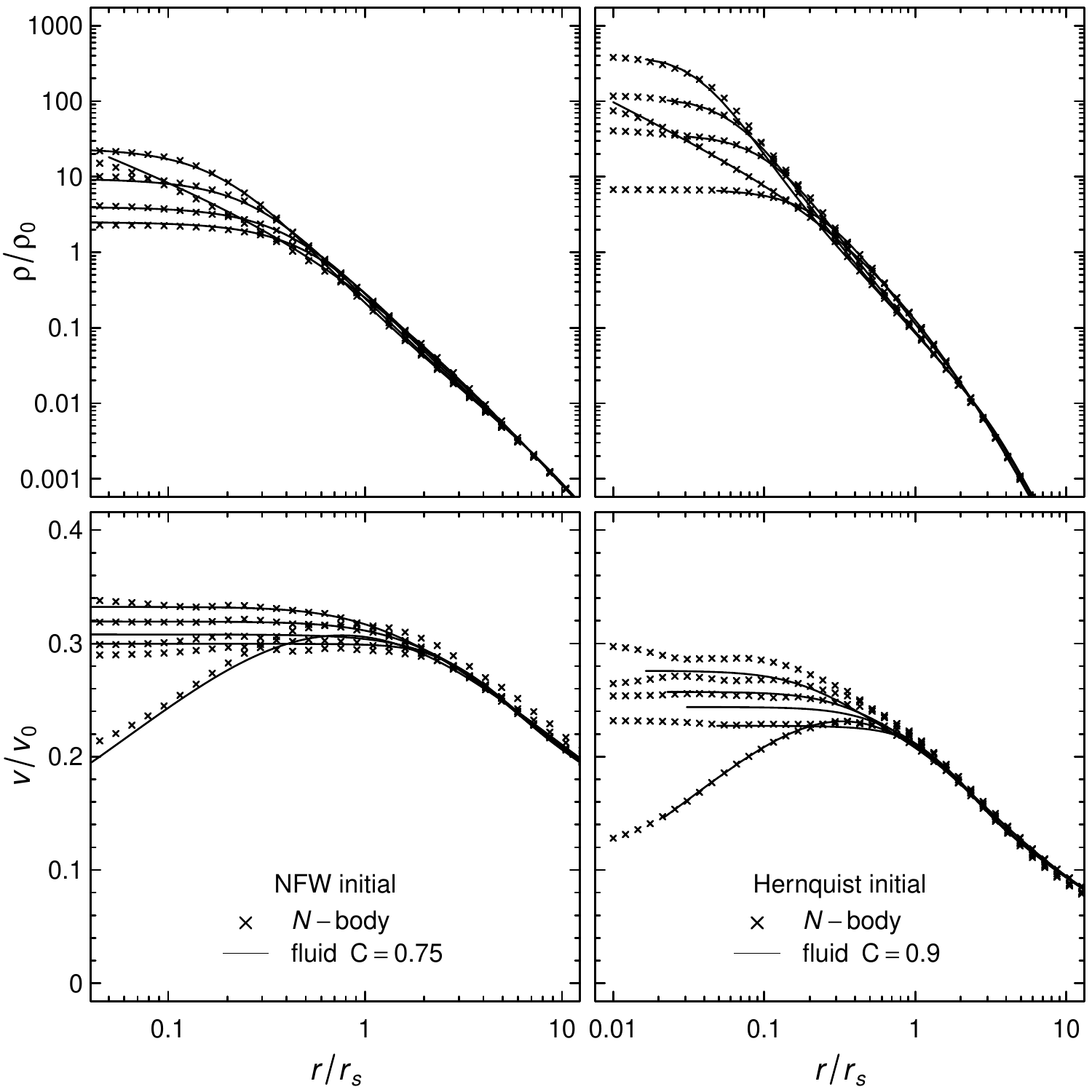}
\caption{Gravothermal collapse of SIDM haloes starting from NFW and
  Hernquist profiles: The density and velocity dispersion
  profiles. (\textit{Left:}) $N$-body snapshots with NFW initial
  condition at $t/t_{r,0} = 0, 197, 336, 462$ and $515$, and profiles
  of the fluid model when they have the same central
  density. (\textit{Right:}) Same as left panels, except with
  Hernquist initial profile, plotted at $t/t_{r,0} = 0, 16, 111, 129$
  and $139$.}
\label{fig-nfw-hqs-profiles}
\end{figure}

\begin{figure}
\includegraphics[width=84mm]{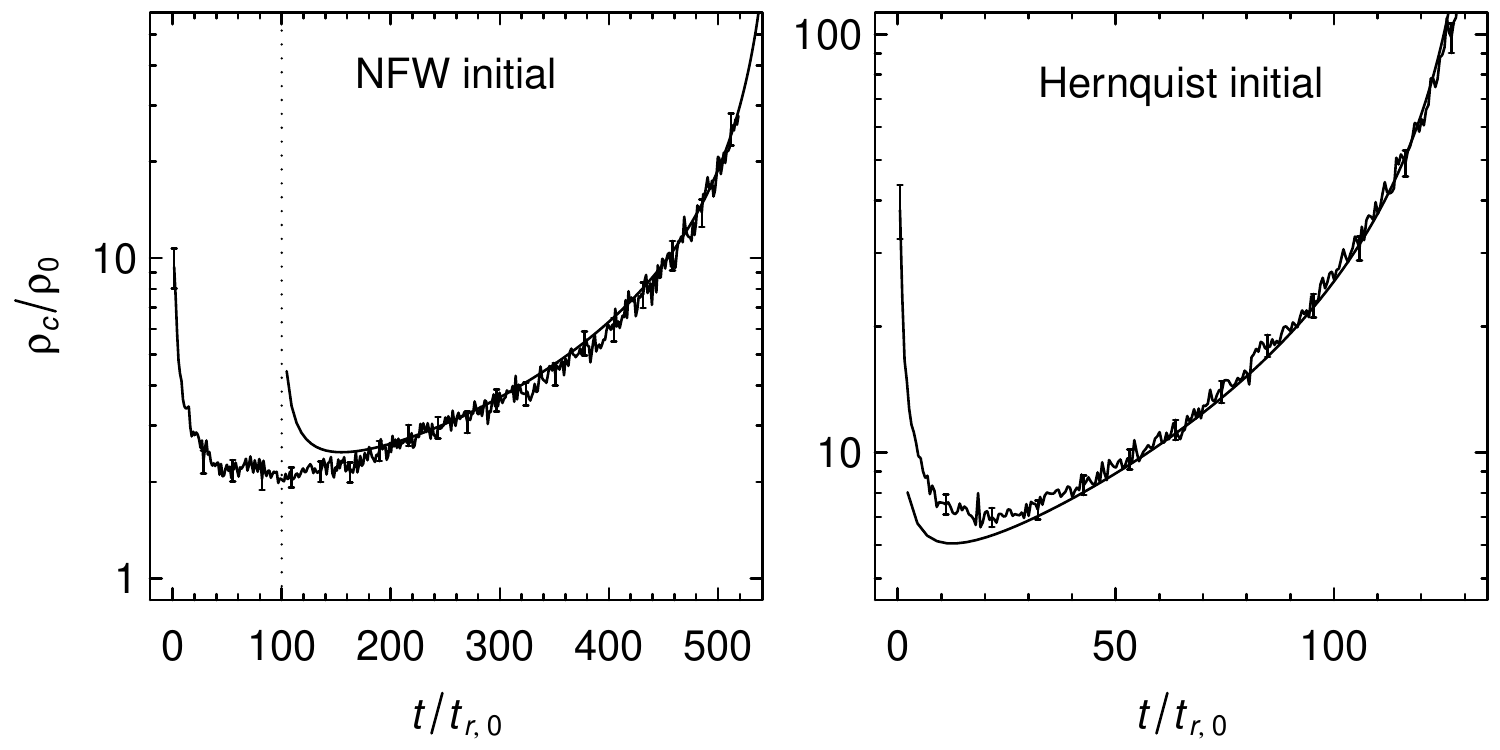}
\caption{Central density versus time, starting from NFW
  (\textit{left}) and Hernquist profiles (\textit{right}).  We use
  $C=0.75$ for NFW and $C=0.9$ for Hernquist profile.  In the left
  panel, the origin for the fluid curve is shifted to the right along
  the time axis by $100\, t_{r,0}$  to show the
  agreement in the gravothermal collapse phase. Minimum density of the
  fluid is $\pm 20$ per cent different from $N$-body. Error bars show
  $\Delta \rho_c / \rho_c = 2/\sqrt{N_c}$ for every 10 data points.  }
\label{fig-nfw-hqs-central}
\end{figure}

Figs.~\ref{fig-nfw-hqs-profiles} and \ref{fig-nfw-hqs-central} show
that the fluid model agrees reasonably with $N$-body results. In
particular, the density profiles in the two cases evolve through a
sequence which is close to the $N$-body results, although the
agreement is better if the following adjustments are made to the time
axis of the fluid model solutions (\ref{fig-nfw-hqs-central}). For
NFW, $N$-body and fluid model central density agree best when compared
at times which differ by a small \textit{shift} equal to
$100\,t_{r,0}$, or $20$ per cent of the collapse time. For the
Hernquist initial profile, the agreement is best if the time axis is
\textit{scaled} by a factor of order of unity which is equivalent to
adjusting the value of $C$ in the conductivity from $C=0.75$ (found in
the BSI run) to $0.9$, which also differs by only $20$ per cent. For
both NFW and Hernquist initial conditions, density profiles agree very
well when the central densities are equal, but the values of the
central densities differ by about $20$ per cent at the maximum core
expansion.

We note that the initial NFW profile evolves very differently
from the initial Hernquist profile, for the same parameters $\rho_0$
and $r_s$. The NFW run has a central density about three times smaller 
at maximum core expansion, and a collapse time about four times longer
than for the H run.  This is because the NFW profile has larger heat
flux at $r \ga r_s$ due to its larger density there, which heats and
expands the central mass more than does the Hernquist profile.

Our simulation results are qualitatively similar to those of
\citet{2000ApJ...543..514K}, but the time evolutions differ by a
factor of two. We do not know the reason for this difference.  We show
how our units convert to those in \citet{2000ApJ...543..514K} as
follows: $\rho_0^{\mbox{\tiny KW}}=2\pi \rho_0$, $t_{r,c}^{\mbox{\tiny
    KW}}=1.7\times 2\sqrt{2} a t_{r,0} = 11 \, t_{r,0}$ and
$\hat{\sigma}_{DM}^{\mbox{\tiny KW}}=2\pi\hat{\sigma}_0$. Their
simulation for $\hat{\sigma}_{DM}^{\mbox{\tiny KW}} = 1$, which has
the same cross section as ours, reached minimum density at $t \approx
t_{r,c}^{\mbox{\tiny KW}} = 11 t_{r,0}$ and collapsed gravothermally
to $\rho_c = 2 \rho_0^{\mbox{\tiny KW}} = 13 \,\rho_0$ at $t \approx
3.2 t_{r,c}^{\mbox{\tiny KW}} = 35 \,t_{r,0}$, while our simulation
reached those densities at $20 \,t_{r,0}$ and $70 \,t_{r,0}$,
respectively (Fig.~\ref{fig-nfw-hqs-central}). Their evolution is
about twice as fast as in our simulation. The origin of this
discrepancy is unknown.

\subsubsection{The Source of Difference}
\label{section-discussion-diff}
Our Monte Carlo $N$-body simulations agree with the solutions of the
conducting fluid model in the self-similar gravothermal collapse
phase, but in general have about a $20$ per cent difference in the
central density and the collapse rate.  This modest disagreement is
not surprising in view of the fact that the conducting fluid model is
not an exact theory derived from first principals.
\citet{1988MNRAS.230..223H} compared the thermal conductivity of the
conducting fluid model with that calculated from the orbit-averaged
Fokker-Planck equation for star clusters with several profiles,
including polytropes and lowered Maxwellians, evaluated at the
centres. They find that the coefficients of conductivity $C$ varied
from one profile to another by factors of 2 or 3. The overall collapse
rates are not that different, probably because the profiles quickly
converge to the self-similar solution around the centre. Indeed, for
star clusters, the value of $C=0.88$, which makes the fluid model
match the asymptotic collapse rate in the isotropic Fokker-Planck
calculation \citep{1980ApJ...242..765C}, also gives the correct
collapse time of the Plummer model, 15.4 half-mass relaxation times
\citep{1987ApJ...313..576G, 1989MNRAS.237..757H}. This means that $C$
needs not to be different for the cases of self-similar collapse
solution and the collapse of Plummer profile.

\begin{figure}
\includegraphics[width=84mm]{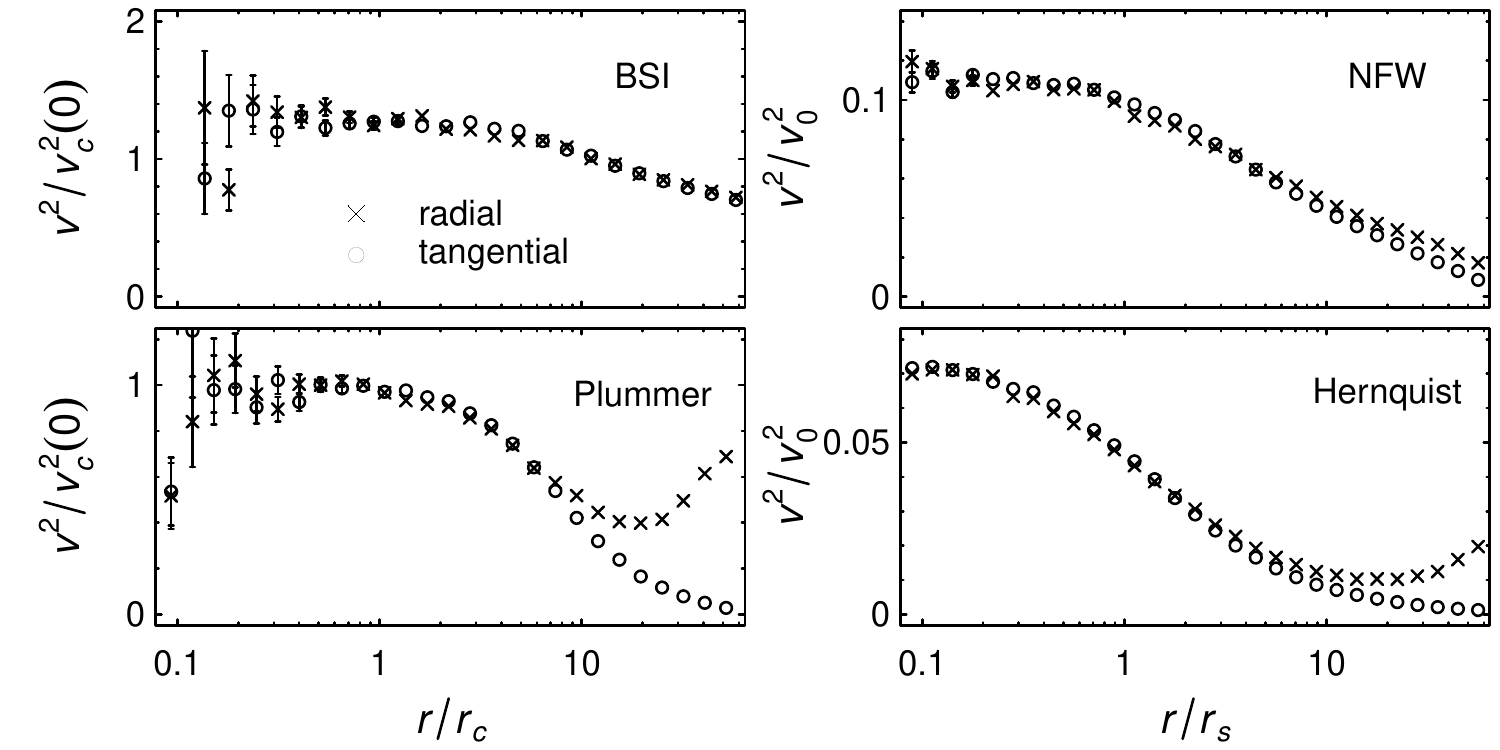}
\caption{Velocity anisotropy for $N$-body results. The radial and
  tangential velocity dispersion for each run. The snapshots are taken
  when the densities at the centre are $10 \rho_c(0)$ for run BSI and
  P, $25 \rho_0$ for run NFW and $125 \rho_0$ for run H. The velocity
  dispersion is calculated in 40 equally-spaced logarithmic bins. The
  error bars show the Poisson fluctuations when they are larger than 1 per
  cent, $\Delta v^2$ = $v^2/\sqrt{N}$, where $N$ is the number of
  particles in each bin.}
\label{fig-anisotropy}
\end{figure} 

One possible distinction between the 1D conducting fluid model and the
3D $N$-body/Monte Carlo simulations is that the former assumes that
particle velocity distributions are isotropic in the frame of the bulk
flow while the latter do not.  Anisotropy in velocity dispersion
affects the collapse time \citep{1983MNRAS.203..811B,
  1990MNRAS.244..478L}. Fokker-Planck calculations with anisotropy
show that the collapse time of the star clusters initialized by the
Plummer model is $20$ per cent larger than for the isotropic case
\citep{1995PASJ...47..561T}, and agrees with $N$-body simulation
\citep{2007MNRAS.374..703K}. In Fig.~\ref{fig-anisotropy}, we plot the
radial ($v_r^2$) and tangential [$v_\perp^2 = (v_\theta^2 +
v_\phi^2)/2$] velocity dispersions of our simulation when the central
density has increased by a factor of 10.  We do not see anisotropies
near the centre, but we do at $r \ga 5r_c$ for runs P, NFW and H.
When particles scatter from the centre to large radii, their nearly
radial orbits bring anisotropy to the initially isotropic velocity
distribution at large radii. If the original unscattered particles at
large radii are less numerous, because the density profile is steeper,
the anisotropy that results from scattering particles from small to
large radii will be relatively larger. This is why anisotropy is
larger if the logarithmic slope of the density profile is steeper.
Anisotropy may therefore play some role in the collapse rate of run P
and H.

In short, the conducting fluid model has been shown to describe the
gravothermal collapse relatively accurately, but we cannot expect very
high precision in general. Therefore, the $20$ per cent match in 
minimum core density of runs NFW and H, and $20$ per cent match in the
gravothermal collapse time for runs P, NFW and H run is a very reasonable
agreement.

\subsection{Transitional Regime}
\label{section-transitional}

When $\mathit{Kn}$ becomes comparable to, or smaller than, one, the
self-similar collapse solution of BSI no longer applies, because of
the presence of the second term in the heat conduction equation
(\ref{eq-bsi_heat}). The time evolution in units of the relaxation
time becomes slower the smaller $\mathit{Kn}$ gets, because smaller
mean free path suppresses particle transport more, which results in
smaller heat conduction \citepalias{2002ApJ...568..475B}.  To test the
heat conduction equation (\ref{eq-bsi_heat}) when both long and short
mean-free-path terms are important (i.e. transitional regime), we
compared the time evolution of Monte Carlo $N$-body simulations with
those of the conducting fluid model, derived numerically when no
self-similar solution exists, with initial Knudsen numbers
$\mathit{Kn}^{-1}(0) = \hat{\sigma}_c(0) = 0.25, 0.5, 0.75$ and $1.0$.
The initial condition is the same as in Section \ref{section-bsi}: the
particle distribution of the BSI self-similar collapse solution. See
Table~\ref{table-transitional} for the summary. $N$-body simulation
with larger $\hat{\sigma}_c$ becomes very difficult because of the
mean free path requirement equation~(\ref{eq-condition_r32}). The
initial ratio of mean free path to the kernel size, $\lambda / r_{32}
\propto \rho^{-2/3}$, at the centre is listed in the table. Our
$\hat{\sigma}_c(0)=1.0$ run violates $\lambda > r_{32}$ at $\rho_c(t)
\sim 8$. The number of particles required to follow the collapse
scales as $\hat{\sigma}_c^3$ beyond this density or cross section.

\begin{table}
\begin{center}
  \begin{tabular}{lcrc}
    \hline
    $\hat{\sigma}_c(0)$ & 
    \centering $N$ & $\lambda/r_{32}$ & $t_{10}/t_{r,c}(0)$ \\
    \hline
    0.25   & $4 \times 64^3$  & 10.3 & 374\\
    0.5    & $4 \times 64^3$  &  5.2 & 417\\
    0.75   & $2 \times 128^3$ &  5.5 & 510\\
    1.0    & $2 \times 128^3$ &  4.1 & 585\\
    \hline
  \end{tabular}
\end{center}
\caption{The transitional regime simulations. $\hat{\sigma}_c$ is the
  initial dimensionless cross section, $N$ is the number of $N$-body
  particles, $\lambda/r_{32}(0)$ is the ratio of the mean free path to
  the kernel size at the centre at $t=0$, and
  $t_{10}$ is the time for the density to increase by a factor of
  10. Other parameters are the same as for the BSI run.  }
  \label{table-transitional}
\end{table}

For the fluid model, the prefactor $C$ is chosen to be $0.75$ to make
an agreement at small $\hat{\sigma}_c$ (Section \ref{section-bsi}). We
ran the fluid code with initial dimensionless cross section
$\hat{\sigma}(0)/\sqrt{b}= 0, 0.25, 0.5, 0.75,1.0,1.5$ and $2.0$. The
time evolution of the fluid model in units of initial relaxation time
depends only on the combination $\hat{\sigma}/\sqrt{b}$, which can be
seen from the heat conduction equation~(\ref{eq-bsi_heat}).

\begin{figure}
\includegraphics[width=84mm]{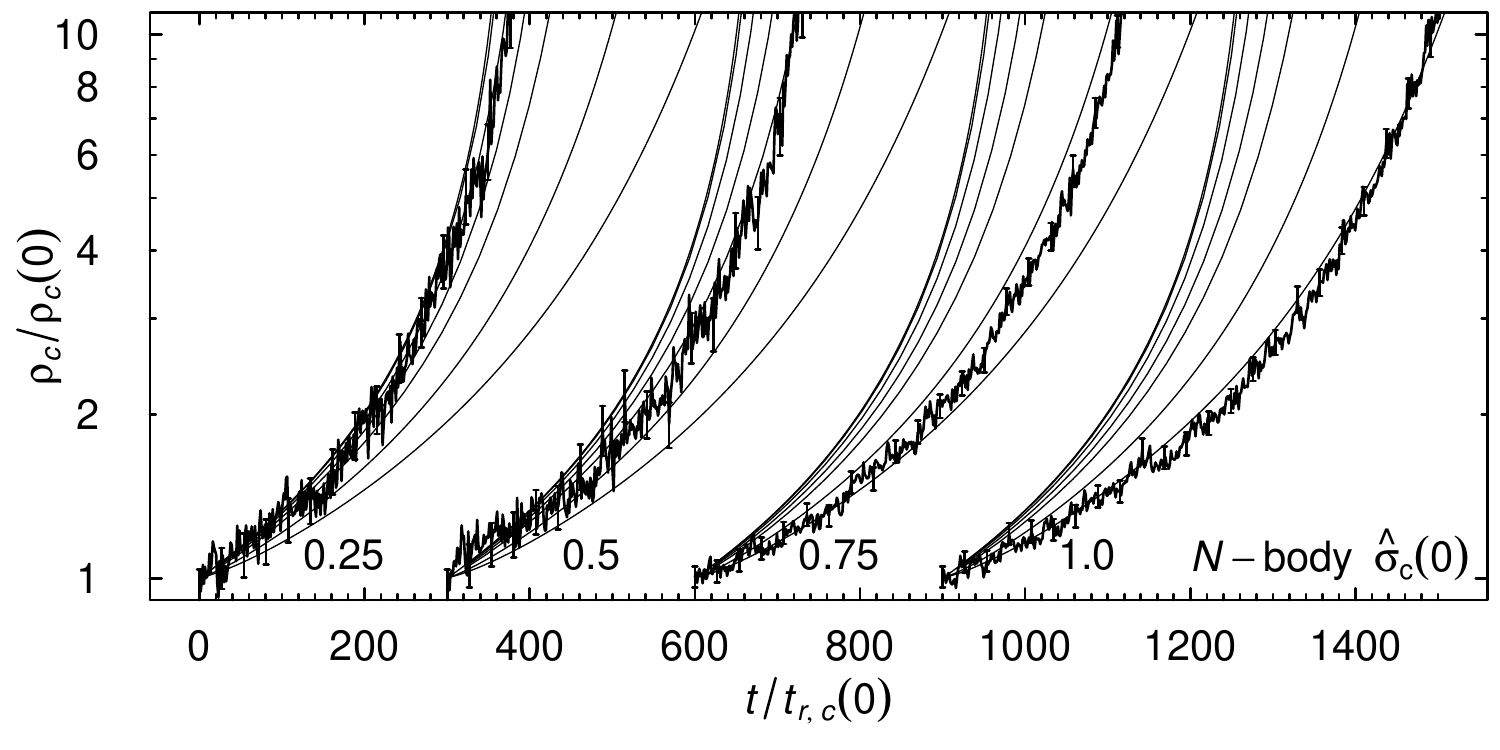}
\caption{Gravothermal collapse in the transitional regime: Central
  density against time from Monte Carlo $N$-body simulations and the
  conducting fluid model.  Cross sections of $N$-body runs are
  $\hat{\sigma}_c(0)=0.25, 0.50, 0.75$ and $1.0$, from left to right,
  shifted to the right by 300 for readability. Four copies of smooth
  curves are the solutions of the fluid model with cross sections
  $\hat{\sigma}_c(0)/\sqrt{b} = 0, 0.25, 0.5, 1.0, 1.5$ and $2.0$,
  from left to right. First two curves, $\hat{\sigma}_c/\sqrt{b} = 0$
  and $0.25$, are indistinguishable, and $\hat{\sigma}_c/\sqrt{b} =
  0.5, 1.0, 0.15$, and $2.0$ overlap with the $N$-body results for
  $\hat{\sigma}_c=0.25, 0.5, 0.75$, and $1.0$, respectively. This match
  suggests $b=0.25$.}
\label{fig-bsi_central_transitional}
\end{figure}

We plot the evolution of the central density for both $N$-body and
fluid model results in Fig.~\ref{fig-bsi_central_transitional}. In
units of the relaxation time, the collapse is slower for larger cross
section.  Fluid model results for $\hat{\sigma}_c(0)/\sqrt{b} = 0.5,
1.0, 1.5$ and $2.0$ evolve similarly to $N$-body simulation results
for $\hat{\sigma}_c(0)= 0.25, 0.5, 0.75$ and $1.0$, respectively. This
suggest that the effective value of the coefficient is $b=0.25$ in the
transitional regime, $\mathit{Kn} \ga 1$.
 
\begin{figure}
\centering
\includegraphics[width=42mm]{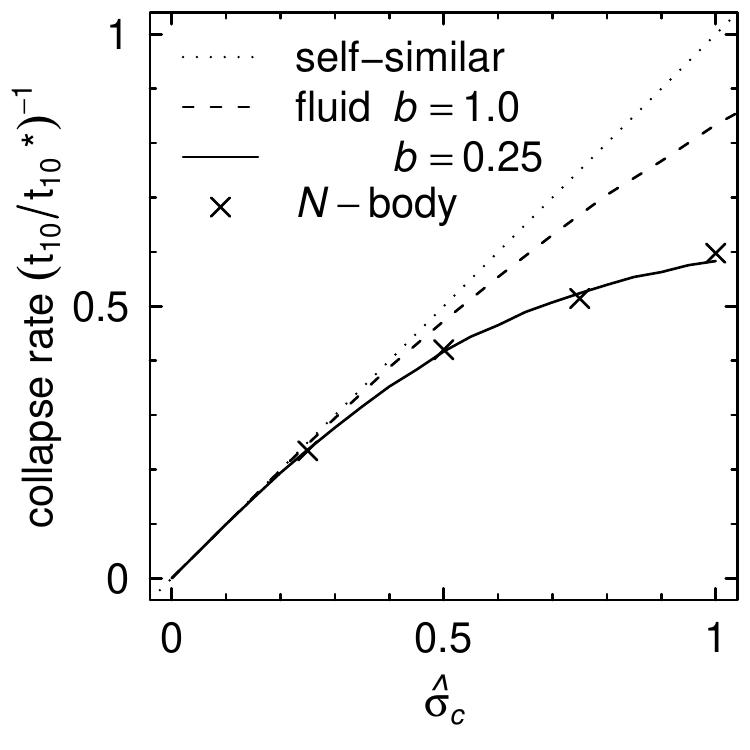}
\caption{ Collapse rate plotted as a function of cross section for
  self-similar solution (dotted line), fluid model with previously
  assumed value $b=1.0$ (dashed line), with $b= 0.25$ (solid line) and
  $N$-body (crosses). See text for the definition of the collapse
  rate.  A value of $b=0.25$ makes the fluid model solution (with its
  conductivity that maps smoothly between the limits of long and short
  mean free path) agree with $N$-body simulations even in the
  transitional regime.}
\label{fig-bsi_collapse_rate}
\end{figure}

To show the agreement between our $N$-body results and the fluid model
solution with $b=0.25$, we plotted the normalized collapse rate as a
function of cross section in Fig.~\ref{fig-bsi_collapse_rate}, where
$t_{10}$ is the time for the density to increase by a factor of 10,
and the fiducial time scale $t_{10}^*$ is the time at which the
self-similar solution with $\hat{\sigma}_c(0)=1$ has a density
increase by a factor of 10. The collapse rate is proportional to the
cross section in the large $\mathit{Kn}$ regime (small
$\hat{\sigma}$), but deviates from the linear relation in the
transitional regime $\hat{\sigma} \ga 0.5$, as predicted by BSI.  The
collapse rate should reach some maximum at some cross section and then
decrease as $t_{10}^{-1} \propto \hat{\sigma}_c^{-1}$ as
$\hat{\sigma}_c \rightarrow \infty$. Furthermore, since the value of
$b$ can be calculated from first principals in the small $\mathit{Kn}$
regime, $b$ should converge to the Chapman-Enskog value (b=1.38 in
Section \ref{section-fluid}) as $\mathit{Kn} \rightarrow 0$.  However,
due to the numerical limit in equation~(\ref{eq-condition_r32}), we
cannot go into the small $\mathit{Kn}$ regime to see the convergence
to the Chapman-Enskog theory or the turn over of the collapse
rate. Our $N$-body results are consistent with a constant $b = 0.25$
in the range we are able to simulate.

\section{Discussion: Implications for the Cosmological Similarity Solution}
\label{section-cosmological-solution}

In this section, we discuss the consequences for the cosmological
similarity solution derived by \citet[hereafter,
A\&S]{2005MNRAS.363.1092A} of replacing the values of the prefactors
(in the heat conduction equation \ref{eq-bsi_heat}) $C=b=1.0$ by
$C=0.75$ and $b=0.25$, as calibrated by our Monte Carlo $N$-body
simulations.  Until now, we have limited our attention to the
gravothermal relaxation of \textit{isolated} SIDM haloes.  However,
the haloes that form in a cosmological context are \textit{not}
isolated but rather build up over time from the nonlinear growth of
small-amplitude initial density perturbations in the expanding
cosmological background universe, which results in a continuous infall
of additional mass. To model this cosmological formation and evolution
of SIDM haloes, \citetalias{2005MNRAS.363.1092A} added the BSI heat
conduction term to the fully time-dependent conservation equations of
the fluid approximation in 1D, spherical symmetry, as described in
\citetalias{2005MNRAS.363.1092A}. They used these equations to derive
an analytical solution for the so-called `secondary infall' problem,
in an Einstein-de Sitter universe, with initial perturbation given by
the spherical overdensity profile $\delta M / \bar{M} \propto
\bar{M}^{-1/6}$, where $\bar{M}(r)$ is the unperturbed mass in a
sphere of radius $r$ at the mean density of the universe, and $\delta
M(r) = M(r) - \bar{M}(r)$ is the mass perturbation inside the radius.
The solution with this initial overdensity is self-similar, that is,
the solution is time-independent if radius and density are measured in
units of the time-varying turn-around radius $r_{\rm ta}$ and
background critical density $\rho_b$, respectively. The family of
similarity solutions is parametrized by the dimensionless cross
section,
\begin{equation}
  Q \equiv \rho_b \sigma r_{\rm vir},
\end{equation}
where $ r_{\rm vir}$ is the halo virial radius, the radius at which an
accretion shock occurs. The density profile has a core whose density
and size depend upon the value of $Q$.

The heat conduction equation~(\ref{eq-bsi_heat}) has a dependence on
$\hat{\sigma}=H/\lambda$ given by,
\begin{equation}
  \label{eq-flux-function}
  L \propto \hat{\sigma} \left( a^{-1} C^{-1} + b^{-1}
    \hat{\sigma}^2 \right)^{-1}.
\end{equation}
This function takes its maximum value $\sqrt{abC}/2$ at $\hat{\sigma}=
\sqrt{a^{-1}b\,C^{-1}}$. Compared to previously assumed values,
$C=b=1.0$, our calibrated heat conduction has a maximum that is
smaller by a factor of $\sqrt{bC}\approx0.4$, which occurs at a value
of $\hat{\sigma}$ that is smaller by a factor of
$\sqrt{bC^{-1}}=0.58$.  There is a minimum central density and largest
core size which occur approximately when the dimensionless cross
section at the centre, $\hat{\sigma}_c$, defined in
Section~\ref{section-definition}, maximizes the heat flux with the
value, $\hat{\sigma}_c = \sqrt{a^{-1} b\,C^{-1}}$.  The solution with
that $\hat{\sigma}_c$ is defined as the maximally-relaxed solution,
and the corresponding cross section $Q$ is denoted by $Q_{\rm th}$
\citepalias{2005MNRAS.363.1092A}. This maximally-relaxed halo has a
density profile almost identical to the empirical Burkert profile
\citep{1995ApJ...447L..25B}, which fits the observed rotation curves
of dwarf and LSB galaxies well.

\begin{figure}
\includegraphics[width=84mm]{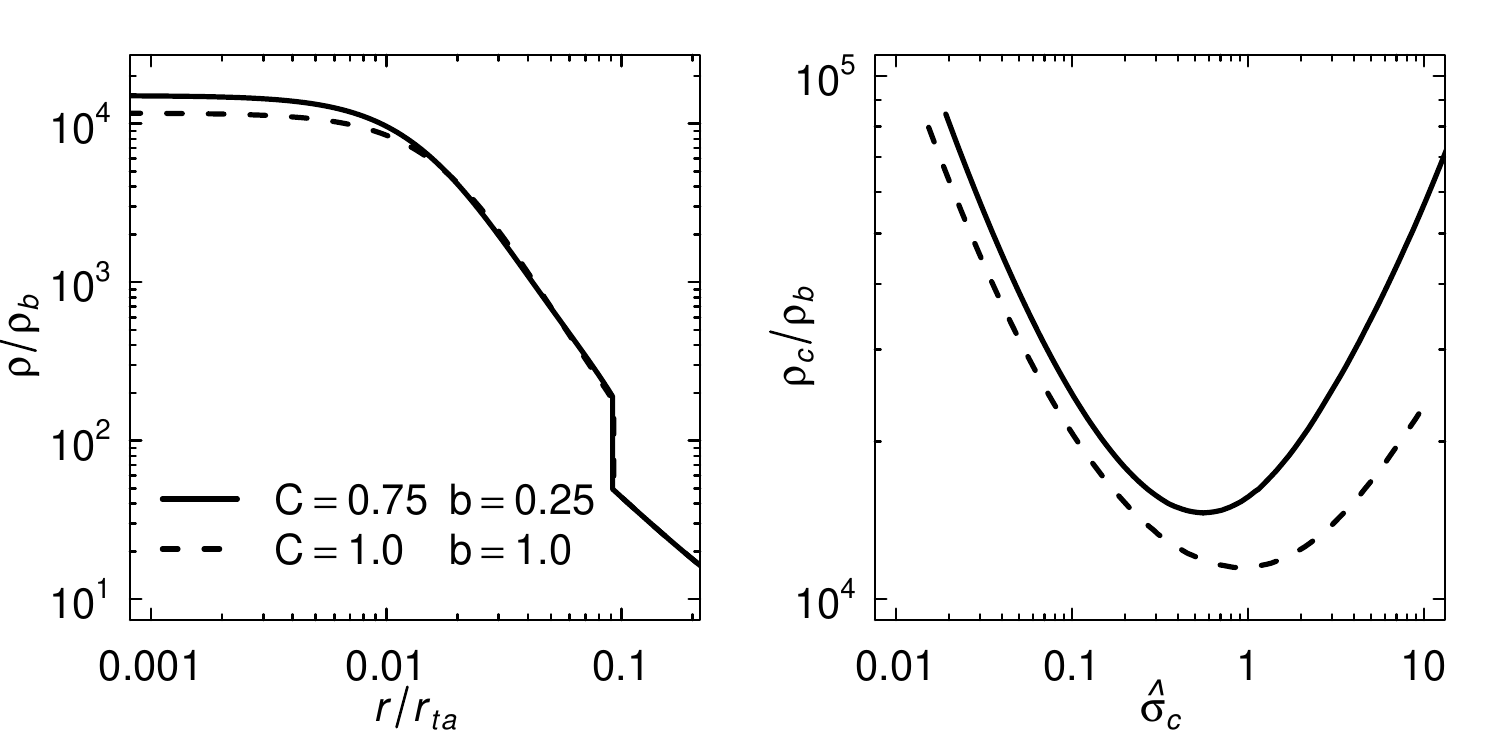}
\caption{The cosmological similarity solution by
  \citet{2005MNRAS.363.1092A} with our calibrated prefactors (C=0.75,
  b=0.25; solid line) and the original prefactors (C=b=1.0; dotted
  line).  (\textit{Left:}) The maximally-relaxed density profiles,
  with density and radius in units of background critical density
  $\rho_b$ and turn around radius $r_{\rm ta}$. The maximally-relaxed
  solution for new prefactors has a 30 per cent larger central
  density.  (\textit{Right:}) The central density for old and new
  prefactors for different cross section values.}
\label{fig-ahn-shapiro}
\end{figure}

We used the same numerical code as \citetalias{2005MNRAS.363.1092A} to
solve for the cosmological similarity solutions with our calibrated
prefactors, and to compare with their original solutions. In
Fig.~\ref{fig-ahn-shapiro}, we plot the maximally-relaxed density
profile (left panel), and the dependence of the central density on
dimensionless cross section $\hat{\sigma}_c$ (right panel).  For
prefactors $C=b=1.0$ adopted by \citetalias{2005MNRAS.363.1092A}, the
solution is maximally relaxed for $Q_{\rm th} = 7.35\times 10^{-4}$,
with central density $\rho_c = 1.17\times 10^4 \rho_b$. For $C=0.75$
and $b=0.25$, the maximally-relaxed solution shifts to $Q_{\rm
  th}=3.95\times 10^{-4}$, with a central density $\rho_c=1.50\times
10^4 \rho_b$.\footnote{The value of $\hat{\sigma}_c$ which defined the
  maximally-relaxed solution, $\hat{\sigma}_c = \sqrt{a^{-1}bC^{-1}}$,
  changes from $0.666$ to $0.384$ if the conductivity parameters
  change their values in A\&S to our new values here.  However, the
  actual minimum densities occur at slightly higher $\hat{\sigma}$.
  For $C=b=1.0$, central density takes its minimum value $\rho_c=1.15
  \times 10^{4} \rho_b$ at $\hat{\sigma}_c=0.85$, or $Q=9.78\times
  10^{-4}$. For $C=0.75$ and $b=0.25$, the minimum density
  $\rho_c=1.46\times 10^4 \rho_b$ is found at $\hat{\sigma}_c=0.55$, or
  $Q=3.95\times 10^{-4}$.}  This central density is about $30$ per
cent larger than that of the original solution.  For $\hat{\sigma}_c
\ll 1$, a cross section that is $1.33$ times as large is required to
achieve the same central density, due to the change of coefficient $C$
from $1.0$ to $0.75$.

\citetalias{2005MNRAS.363.1092A} estimated the cross section values
that brought dwarf galaxy rotation curves into agreement with the
cosmological self-similar halo profiles. In order to translate a given
value of $Q_{\rm th}$ into a cross section value $\sigma$, the typical
formation time for haloes that host dwarf galaxies in the $\Lambda$CDM
model of structure formation was used to relate $r_{\rm vir}$ and
$\rho_b$ in the definition of $Q$, as described in
\citetalias{2005MNRAS.363.1092A}.  Using the same argument, our
corrected $Q_{\rm th}$ value corresponds to a cross section $117
\textrm{ cm}^2 \textrm{ g}^{-1}$, which is smaller than the original
value $218 \textrm{ cm}^2 \textrm{ g}^{-1}$ but still much larger than
the values, $0.5-5 \textrm{ cm}^2 \textrm{ g}^{-1}$, found in Monte
Carlo $N$-body simulations that produce cored profiles for galactic
haloes formed from cosmological initial conditions
\citep{2001ApJ...547..574D}.

We will discuss this apparent discrepancy between the values of the
SIDM cross section which are best able to match the observed rotation
curves of dwarf and LSB galaxies, from the A\&S similarity solutions
and cosmological $N$-body/Monte Carlo simulations, respectively, in a
separate paper. Here, we have demonstrated that this discrepancy is
not likely to result from the break down of the conducting fluid
model. We have found good agreement, in fact, between the
Monte Carlo $N$-body simulations and the conducting fluid model for
the thermal relaxation and gravothermal collapse of \textit{isolated}
haloes (of fixed mass), at least with regard to the long
mean-free-path regime and the transitional regime in which the mean
free path is comparable to the system size. These are the regimes of
greatest relevance to the SIDM halo problem. We have also shown that,
when we use the agreement between our Monte Carlo $N$-body results and
the conducting fluid model solutions for \textit{isolated} haloes to
calculate the dimensionless parameters on which the heat conduction
depends, the impact of the modified parameters on the A\&S similarity
solution is relatively small. We must, therefore, seek a different
explanation for the different values of $\sigma$ required by the
cosmological $N$-body/Monte Carlo and the similarity solutions of
\citetalias{2005MNRAS.363.1092A}, to produce the same observed degree
of relaxation of the SIDM halo density profiles.

As we shall discuss in our next paper in this series, the essential
difference between the self-similar solution of
\citetalias{2005MNRAS.363.1092A} and the cosmological $N$-body
simulations is that the self-similar halo is continuously heated by
the accretion shock, while the inner part of a more realistic halo, on
the galaxy scale, is eventually unaffected by infall when infall rate
slows after the halo forms.
We will compare cosmological Monte Carlo $N$-body simulations with the
conducting fluid model with non-self-similar infall in our next paper.

\section{Conclusion and Summary}
\label{section-conclusion}
The ability of the SIDM hypothesis to resolve the cusp-core problem of
collisionless CDM haloes on the galaxy scale and the cross section
values required to accomplish this remain uncertain, as long as the
previous conclusions drawn from $N$-body simulation and 
the conducting fluid model differ so strongly. To rectify
this situation, we have developed a new Monte Carlo $N$-body code of
our own, based on the pre-existing \textsc{GADGET} 1.1 $N$-body code,
and applied it to compare the $N$-body simulations with
the conducting fluid model for isolated, spherically symmetric
self-gravitating SIDM haloes.  The collisions were assumed to be
velocity independent, elastic, and isotropic. Our results include the following:

\begin{enumerate}
\item{Our Monte Carlo $N$-body simulations are in very good agreement
    with the analytical self-similar gravothermal collapse solution of
    \citetalias{2002ApJ...568..475B}, when the coefficient of
    thermal conduction is set to $C = 0.75$ (Section
    \ref{section-bsi}); the density and velocity dispersion profiles
    evolve self-similarly; the central density and velocity dispersion
    follow the self-similar time evolution formulae with the predicted
    constant $\alpha=2.19$; and, the time to collapse is always
    proportional to the central relaxation time at that time (equation
    \ref{eq-self_similar_1}-\ref{eq-self_similar_tcoll}).  }

\item{The conducting fluid model agrees with Monte Carlo $N$-body
    simulations reasonably well for different initial conditions:
    Plummer model, Hernquist profile, and NFW profile, in the
    large-$\mathit{Kn}$ regime. The collapse time and the central
    density at maximum core expansion agree within $20$ per cent. The
    shape of the density profile and the central density evolution as
    a function of time during gravothermal collapse agree very well.}

\item{We also showed that the collapse time becomes longer in units of
    relaxation time as the system transitions from the large- to the
    small-$\mathit{Kn}$ regime, as predicted by the conducting fluid
    model.  The $N$-body results agree with the conducting fluid model
    for $\mathit{Kn} \ge 1$, or $\hat{\sigma} \le 1$, with the
    prefactor $b=0.25$. However, this prefactor is more than five
    times smaller than the Chapman-Enskog value, valid asymptotically
    in the small $\mathit{Kn}$ limit. The conducting fluid model must
    be further calibrated against $N$-body simulations if it is used
    beyond the transitional regime, for $\hat{\sigma} \ga 1$.  }

\item{Our calibration of the prefactors $C$ and $b$ does not change
    the cosmological similarity solutions of
    \citet{2005MNRAS.363.1092A} significantly. The cross section that
    gives the minimum central density on the dwarf galaxy scale is
    altered from $220 \textrm{ cm}^2 \mathrm{g}^{-1}$ to $117 \textrm{
      cm}^2 \mathrm{g}^{-1}$, but this is still much larger than the
    values that cosmological Monte Carlo $N$-body simulations used to
    make cored SIDM haloes ($\sigma \sim 0.5-5 \textrm{ cm}^2 \mathrm{
      g}^{-1}$).  We will investigate this problem in our subsequent
    paper.  Our results here suggest that this apparent discrepancy is
    not the result of a breakdown of either the conducting fluid
    model or the Monte Carlo scattering algorithm in the $N$-body
    simulations. As we shall show in a companion paper, in fact, the
    discrepancy results, instead, from the gradual departure of halo
    evolution from self-similarity as the infall rate during
    cosmological structure formation drops below the self-similar rate
    at late times after halo formation.}

\end{enumerate}

\section{Acknowledgments}
We thank Kyungjin Ahn for useful discussions and letting us use the
cosmological similarity solution solver. This work was supported in
part by NASA ATP grants NNG04G177G and NNX07AH09G, CHANDRA SAO grant
TM8-9009X, and NSF grant AST-0708176 and AST-1009799 to
PRS. Simulations were performed at the Texas Advanced Computing
Center.

\bibliographystyle{mn2e}
\bibliography{junkoda}

\label{lastpage}

\end{document}